\documentclass[11pt, peereview, onecolumn, letterpaper]{IEEEtran}

\usepackage{color}
\usepackage{bbm}
\usepackage{amsfonts}
\usepackage{epsfig}
\usepackage{mathtools}
\usepackage{cite}
\usepackage{setspace}
\usepackage{amsthm}
\usepackage[font=footnotesize]{caption}
\usepackage{subcaption}
\usepackage{slashbox}
\usepackage{tikz}
\usetikzlibrary{matrix}

\date{}

\renewcommand{\IEEEQED}{\QEDopen}
\newcommand{\eqdef}{\stackrel{\triangle}{=}}
\renewcommand{\thesection}{\Roman{section}}
\renewcommand{\thesubsection}{\Alph{subsection}}
\renewcommand{\thesubsubsection}{\alph{subsubsection}}
\DeclareMathOperator*{\argmax}{arg\,max}
\DeclareMathOperator*{\argmin}{arg\,min}
\def\multiset#1#2{\ensuremath{\bigg(\kern-.4em\bigg(\genfrac{}{}{0pt}{}{#1}{#2}\bigg)\kern-.4em\bigg)}}
\newtheoremstyle{mynewtheorem}
{3pt}
{3pt}
{}
{}
{\itshape}
{:}
{.5em}
{\thmname{#1}\thmnumber{ #2}}%

\theoremstyle{mynewtheorem}
\newtheorem{myprop}{Proposition}

\renewcommand*{\arraystretch}{0.5}

\title{\hfill{\small{\it Submitted to IEEE Transactions on Signal Processing - Manuscript ID: T-SP-15743-2013}}\\
\vspace{1cm}\renewcommand{\baselinestretch}{1}\Huge Optimal Algorithms for $L_1$-subspace Signal Processing%
\thanks{This paper was presented in part at the Tenth International Symposium on Wireless Communication Systems (ISWCS), Ilmenau, Germany, August 2013.}}

\renewcommand{\baselinestretch}{2}

\author{
\vspace{-0.3cm}
\IEEEauthorblockN{Panos P. Markopoulos$^\dagger$, George N. Karystinos$^\ddagger$, and Dimitris A. Pados$^{\dagger *}$\footnote{$^*$Corresponding author.}}\\
\IEEEauthorblockA{
\vspace{0.2cm}
\vspace{-0.2 cm} $^\dagger$Department of Electrical Engineering\\
\vspace{-0.5 cm} The State University of New York at Buffalo\\
\vspace{-0.5  cm} Buffalo, NY 14260 USA\\ 
\vspace{-0.5 cm} E-mail: \texttt{\{pmarkopo, pados\}@buffalo.edu} \\
\vspace{0.2cm}
 \vspace{-0.2 cm}  $^\ddagger$Department of Electronic and Computer Engineering\\
\vspace{-0.5 cm} Technical University of Crete\\
\vspace{-0.5  cm} Chania, 73100 Greece \\
\vspace{-0.5 cm} E-mail: \texttt{karystinos@telecom.tuc.gr} \\ 
 }
EDICS: MLR-ICAN, MLR-LEAR, MDS-ALGO, ASP-ANAL, SSP-SSAN\\
 \vspace{-0.1cm}
 Submitted: June 27, 2013 - Revised: April 4, 2014
}

\begin{document}

\maketitle

\thispagestyle{empty}

\vspace{-1.8cm} 
\renewcommand{\baselinestretch}{1.4}

\begin{abstract}
\vspace{-0.2cm} 
We describe ways to define and calculate $L_1$-norm signal subspaces which are less sensitive to outlying data than $L_2$-calculated subspaces.
We start with the computation of the $L_1$ maximum-projection principal component of a data matrix containing $N$ signal samples of dimension $D$.
We show that while the general problem is formally NP-hard in asymptotically large $N$, $D$, the case of engineering interest of fixed dimension $D$ and asymptotically large sample size $N$ is not.
In particular, for the case where the sample size is less than the fixed dimension ($N<D$), we present in explicit form an optimal algorithm of computational cost $2^N$.
For the case $N \geq D$, we present an optimal algorithm of complexity $\mathcal O(N^D)$.
We generalize to multiple $L_1$-max-projection components and present an explicit optimal $L_1$ subspace calculation algorithm of complexity
${\mathcal O}(N^{DK-K+1})$
where $K$ is the desired number of $L_1$ principal components (subspace rank).
We conclude with illustrations of $L_1$-subspace signal processing in the   fields of data dimensionality reduction, direction-of-arrival estimation, and image conditioning/restoration.
\end{abstract}

\vspace{-0.1cm} 
 

\renewcommand{\baselinestretch}{2}

\normalsize
\setcounter{page}{1}

\section{Introduction}
A general intention of subspace signal processing is to partition the vector space of the observed data and isolate the subspace of the signal component(s) of interest from the disturbance (noise) subspace.
Subspace signal processing theory and practice rely, conventionally, on the familiar $L_2$-norm based singular-value decomposition (SVD) of the data matrix.
The SVD solution traces its origin to the fundamental problem of $L_2$-norm low-rank matrix approximation \cite{Eckart1936}, which is equivalent to the problem of maximum $L_2$-norm data projection with as many projection (``principal'') components as the desired low-rank value \cite{Golub1996}.
Among the many strengths of $L_2$-norm principal component analysis (PCA), one may point out the simplicity of the solution, scalability (new principal directions add on to the previous ones), and correspondence to maximum-likelihood estimation (MLE) under the assumption of  additively Gaussian-noise corrupted data.

Practitioners have long observed, however, that $L_2$-norm PCA is sensitive to the presence of outlier values in the data matrix, that is, erroneous values that are away from the nominal data, appear only few times in the data matrix, and are not to appear again under normal system operation upon design.
Recently, there has been an --arguably small but growing-- interest in pursuing $L_1$-norm based approaches to deal with the problem of outliers in principal-components design [\citen{Ke2003}]-[\citen{Li2010}].%
\footnote{Absolute-value errors put significantly less emphasis on extreme errors than squared-error expressions.}
The growth in interest can also be credited incidentally to the popularity of compressed sensing methods [\citen{Donoho2006}]-[\citen{Gao}] that rely on $L_1$-based calculations in signal reconstruction.

This paper makes a case for $L_1$-subspace signal processing.
Interestingly, in contrast to $L_2$, subspace decomposition under the $L_1$ error minimization criterion and the $L_1$ projection maximization criterion are not the same.
A line of recent research pursues calculation of $L_1$ principal components under error minimization 
[\citen{Ke2003}]-[\citen{Brooks2013b}].
The error surface is non-smooth and the problem non-convex resisting attempts to guaranteed optimization even with exponential computational cost.
Suboptimal algorithms may be developed by viewing the minimization function as a convex nondifferentiable function with a bounded Lipschitz constant \cite{Combettes2002}, \cite{Nesterov2013}.
A different approach is to calculate subspace components by $L_1$ projection maximization [\citen{Galpin1987}]-[\citen{Gu2012}].%
\footnote{A combined $L_1$/$L_2$-norm approach has been followed in  [\citen{Ding2006}], [\citen{Li2010}].}
No algorithm has appeared so far with guaranteed convergence to the criterion-optimal subspace and no upper bounds are known on the expended computational effort.

In this present work, given any data matrix ${\mathbf X}\in \mathbb R^{D \times N}$ of $N$ signal samples of dimension $D$, we show that the general problem of finding the maximum $L_1$-projection principal component of $\mathbf X$ is formally NP-hard for asymptotically large $N$, $D$.
We prove, however, that the case of engineering interest of   fixed given dimension $D$ is not NP-hard.
In particular, for the case where $N<D$, we present in explicit form an algorithm to find the optimal component with computational cost $2^N$.
For the case where the sample size exceeds the data dimension ($N \geq D$) --which is arguably of higher interest in signal processing applications-- we present an algorithm that computes the $L_1$-optimal principal component with complexity $\mathcal O\big(N^{\text{rank}(\mathbf X)}\big)$,  $\text{rank}(\mathbf X) \leq D$.
We generalize the effort to the problem of calculating $K$, $1<K \leq \text{rank}( \mathbf X)$, $L_1$ components (necessarily a joint computational problem) and present an explicit optimal algorithm for multi-component subspace design of complexity ${\mathcal O}(N^{\text{rank}(\mathbf X)K-K+1})$.
We conclude with illustrations of the developed $L_1$ subspaces in problems from the fields of dimensionality reduction, direction-of-arrival estimation, and image reconstruction that demonstrate the inherent outlier resistance of $L_1$ subspace signal processing.

The rest of the paper is organized as follows.
Section II presents the problem statement and establishes notation.
Section III is devoted to the optimal computation of the $L_1$ principal component.
Section IV generalizes to optimal $L_1$-subspace calculation (joint multiple $L_1$ components).
Experimental illustrations are given in Section V and a few concluding remarks are drawn in Section VI.

\section{Problem Statement}

Consider $N$ real-valued measurements $\mathbf x_1, \mathbf x_2, \ldots, \mathbf x_N$ of dimension $D$ that form the $D\times N$ data matrix
\begin{equation}
{\bf X}=[{\bf x}_1\;\;{\bf x}_2\;\ldots\;{\bf x}_N].
\end{equation}
In the common version of the low-rank approximation problem, one seeks to describe (approximate)  data matrix ${\bf X}$ by a rank-$K$ product ${\bf R}{\bf S}^T$ where ${\bf R} \in \mathbb R^{D \times K}$, ${\bf S} \in \mathbb R^{N \times K}$, $K \leq\min(D,N)$.
Given the observation data matrix ${\bf X}$, $L_2$-norm matrix approximation minimizes the sum of the element-wise squared error between the original matrix and its rank-$K$ surrogate  in the form of Problem $\mathcal P_1^{L_2}$  defined below,
\begin{equation}
\begin{split}
{\mathcal P}^{L_2}_1:\;\;\;\;&\left({\bf R}_{L_2},{\bf S}_{L_2}\right)=\argmin_{{\bf R}\in{\mathbb R}^{D\times K},\;{\bf S}\in{\mathbb R}^{N\times K}}\left\|{\bf X}-{\bf R}{\bf S}^T\right\|_2
\end{split}
\label{eq:RS}
\end{equation}
where $\|{\bf A}\|_2=\sqrt{\sum_{i,j}|A_{i,j}|^2}$ is the $L_2$ matrix norm (that is, Frobenius norm) of a matrix $\mathbf A$ with elements $A_{i,j}$.
Problem ${\mathcal P}^{L_2}_1$ is our most familiar $K$-singular-value-decomposition ($K$-SVD) problem solved with computational complexity $\mathcal O \big((D+N)\min^2(D,N)\big)$ \cite{Golub1996}. 
$\mathcal P_1^{L_2}$ corresponds also to the statistical problem of maximum-likelihood estimation (MLE) of an unknown rank-$K$ matrix corrupted by additive element-wise independent Gaussian noise~\cite{Vantrees}.

We may expand~(\ref{eq:RS}) to $\displaystyle\min_{{\bf R}\in{\mathbb R}^{D\times K}}\min_{{\bf S}\in{\mathbb R}^{N\times K}}\left\|{\bf X}-{\bf R}{\bf S}^T\right\|_2$ and  inner minimization results to ${\bf S}={\bf X}^T{\bf R}$ for any fixed ${\bf R}$, $\mathbf R^T \mathbf R = \mathbf I_K$, by the Projection Theorem \cite{Golub1996}.
Hence, we obtain the equivalent problem
\begin{equation}
\begin{split}
{\mathcal P}^{L_2}_2:\;\;\;\;&{\bf R}_{L_2}=\argmin_{
{\bf R}\in{\mathbb R}^{D\times K},\,{\bf R}^T{\bf R}={\bf I}_K}\left\|{\bf X}-{\bf R}{\bf R}^T{\bf X}\right\|_2
\end{split}
\label{eq:RR}
\end{equation}
frequently referred to as left-side $K$-SVD.
Since $\left\|{\bf A}\right\|_2^2=\text{tr}\left({\bf A}^T{\bf A}\right)$ where $\text{tr}(\cdot)$ denotes the trace of a matrix, the $L_2$ error minimization problem ${\mathcal P}^{L_2}_2$ is also equivalent to the $L_2$ projection (energy) maximization problem
\begin{equation}
\begin{split}
{\mathcal P}^{L_2}_3:\;\;\;\;&{\bf R}_{L_2}=\argmax_{{\bf R}\in{\mathbb R}^{D\times K},\,{\bf R}^T{\bf R}={\bf I}_K}\left\|{\bf X}^T{\bf R}\right\|_2.
\end{split}
\label{eq:R}
\end{equation}
The optimal  ${\bf R}_{L_2}$ (in ${\mathcal P}^{L_2}_1$, ${\mathcal P}^{L_2}_2$, and ${\mathcal P}^{L_2}_3$) is  known  simply as the $K$ dominant-singular-value left singular vectors of the original data matrix or $K$ dominant-eigenvalue eigenvectors of $\mathbf X \mathbf X^T$~\cite{Eckart1936},~\cite{Golub1996}.
Note that, if $K<D$ and we possess the solution ${\bf R}_{L_2}^{(K)}$ for $K$ singular/eigen vectors  in (\ref{eq:RS}), (\ref{eq:RR}), (\ref{eq:R}), then the solution for rank $K+1$ is derived readily by ${\bf R}_{L_2}^{(K+1)}=\left[{\bf R}_{L_2}^{(K)}\;\;{\bf r}_{L_2}^{(K+1)}\right]$ with
\begin{equation}
{\bf r}_{L_2}^{(K+1)}=\argmax_{{\bf r}\in{\mathbb R}^D,\;\left\|{\bf r}\right\|_2=1}\left\|{\bf X}^T \left({\bf I}_D-{\bf R}_{L_2}^{(K)}{{\bf R}_{L_2}^{(K)}}^T\right){\bf r} \right\|_2.
\end{equation}
This is known as the PCA scalability property.

$L_2$ PCA, as reviewed above in ${\mathcal P}^{L_2}_1$, ${\mathcal P}^{L_2}_2$, and ${\mathcal P}^{L_2}_3$, has a simple solution, is scalable (new principal directions add on to the previous ones), and corresponds to MLE under the assumption of Gaussian additively corrupted data.
Practitioners, however, have long noticed a drawback.
By minimizing the sum of squared errors, $L_2$ principal component calculation becomes sensitive to extreme error value occurrences caused by  the presence of outlier measurements in the data matrix (measurements that are numerically distant from the nominal data, appear only few times in the data matrix, and are not to appear under normal system operation upon design).
Motivated by this observed drawback of $L_2$ subspace signal processing, in this work we study and  pursue subspace-decomposition approaches that are based on the $L_1$ norm,
\begin{equation}
\left\|{\bf A}\right\|_1=\sum_{i,j}\left|A_{i,j}\right|.
\end{equation}
We  may ``translate" the three equivalent $L_2$ optimization problems (\ref{eq:RS}), (\ref{eq:RR}), (\ref{eq:R}) to new problems that utilize the $L_1$ norm as follows,
\begin{align}
\begin{split}
&{\mathcal P}^{L_1}_1:\;\;\;\;\left({\bf R}_{L_1},{\bf S}_{L_1}\right)=\argmin_{{\bf R}\in{\mathbb R}^{D\times K},\;{\bf R}^T{\bf R}={\bf I}_K,\;{\bf S}\in{\mathbb R}^{N\times K}}\left\|{\bf X}-{\bf R}{\bf S}^T\right\|_1, 
\end{split} \\
\label{eq:RSL1}
\begin{split}
&{\mathcal P}^{L_1}_2:\;\;\;\;{\bf R}_{L_1}=\argmin_{{\bf R}\in{\mathbb R}^{D\times K},\;  {\bf R}^T{\bf R}={\bf I}_K}\left\|{\bf X}-{\bf R}{\bf R}^T{\bf X}\right\|_1, 
\end{split} \\
 \begin{split}
&{\mathcal P}^{L_1}_3:\;\;\;\;{\bf R}_{L_1}=\argmax_{{\bf R}\in{\mathbb R}^{D\times K}, \;  {\bf R}^T{\bf R}={\bf I}_K}\left\|{\bf X}^T{\bf R}\right\|_1. 
\end{split}
\label{eq:RL1}
\end{align}
A few comments appear useful at this point: 
(i) ${\mathcal P}^{L_1}_1$ corresponds to MLE when the additive noise disturbance follows a Laplacian distribution \cite{Vantrees}.
(ii) The optimal metric value in ${\mathcal P}^{L_1}_3$ with a single dimension ($K=1$) is the complexity parameter for saddle-point methods when used to provide an approximate solution to the $\ell_1$/nuclear-norm Dantzig selector problem \cite{Nesterov2013}.
(iii) Under the $L_1$ norm, the three optimization problems ${\mathcal P}^{L_1}_1$, ${\mathcal P}^{L_1}_2$, and ${\mathcal P}^{L_1}_3$ are no longer equivalent.
(iv) Under $L_1$, the PCA scalability property does not hold (due to loss of the Projection Theorem).
(v) Even for reduction to a single dimension (rank $K=1$ approximation), the three problems are difficult to solve.
(vi) As of today, it is unknown which of the  subspaces defined in $\mathcal P_{1}^{L_1}$, $\mathcal P_{2}^{L_1}$, and $\mathcal P_{3}^{L_1}$ exhibits stronger resistance against faulty measurements; indeed, none of these problems had been solved optimally so far for general $D,K$.

In this present work, we focus exclusively on ${\mathcal P}^{L_1}_3$.
In Section III, we seek to find efficiently the principal maximum $L_1$  projection component of ${\bf X}$.
In Section IV, we investigate the problem of calculating (jointly necessarily) multiple ($K>1$) $L_1$ projection components that maximize the $L_1$ ``energy" of the data on the projection subspace.

\section{The $L_1$-norm Principal Component}
\label{sec:OneComponent}

In this section, we concentrate on the calculation of the $L_1$-maximum-projection component of a data matrix $\mathbf X \in \mathbb R^{D \times N}$ (Problem $\mathcal P_3^{L_1}$ in (\ref{eq:RL1}), $K=1$).
First, we  show that the problem is in general NP-hard and   review briefly suboptimal techniques from the literature.
Then, we  prove that, if the data dimension $D$ is fixed, the principal $L_1$-norm component is in fact computable in polynomial time and present an algorithm that calculates the $L_1$ principal component of $\mathbf X$ with complexity ${\mathcal O}\left(N^{\text{rank}({\bf X})}\right)$, $\text{rank}({\bf X})\leq D$.

\subsection{Hardness of Problem and an Exhaustive-search Algorithm Over the Binary Field}

We present a fundamental property of Problem  ${\mathcal P}^{L_1}_3$, $K=1$, that will lead us to an efficient solution.
The property is presented in the form of Proposition \ref{prop:quad} below and interprets ${\mathcal P}^{L_1}_3$ as an equivalent quadratic-form maximization problem over the binary field.
\begin{myprop}
For any data matrix $\mathbf X \in \mathbb R^{D \times N}$, the solution to 
${\mathcal P}^{L_1}_3:{\bf r}_{L_1}=\argmax_{{\bf r}\in{\mathbb R}^D,\left\|{\bf r}\right\|_2=1}\left\|{\bf X}^T{\bf r}\right\|_1$
is given by
\begin{equation}
{\bf r}_{L_1}=\frac{{\bf X}{\bf b}_\text{opt}}{\left\|{\bf X}{\bf b}_\text{opt}\right\|_2}
\label{eq:rL1}
\end{equation}
where\begin{equation}
{\bf b}_\text{opt}=\argmax_{{\bf b}\in\{\pm1\}^N}\left\|{\bf X}{\bf b}\right\|_2=\argmax_{{\bf b}\in\{\pm1\}^N}{\bf b}^T{\bf X}^T{\bf X}{\bf b}.
\label{eq:bopt}
\end{equation}
In addition,
$\left\|{\bf X}^T {\bf r}_{L_1}\right\|_1=\left\|{\bf X}{\bf b}_\text{{opt}}\right\|_2$.
\label{prop:quad}
\end{myprop}

\noindent \emph{Proof:}
For any ${\bf z}\in{\mathbbm R}^N$, $\displaystyle\left\|{\bf z}\right\|_1=\text{sgn}\left({\bf z}\right)^T{\bf z}=\max_{{\bf b}\in\{\pm1\}^N}{\bf b}^T{\bf z}$.
Therefore,  we can rewrite the optimization problem as
\begin{equation}
\max_{\left\|{\bf r}\right\|_2=1}\left\|{\bf X}^T{\bf r}\right\|_1=\max_{\left\|{\bf r}\right\|_2=1}\max_{{\bf b}\in\{\pm1\}^N}{\bf b}^T{\bf X}^T{\bf r}=\max_{{\bf b}\in\{\pm1\}^N}\max_{\left\|{\bf r}\right\|_2=1}{\bf r}^T{\bf X}{\bf b}.
\label{eq:rXb}
\end{equation}
For any fixed vector ${\bf b}$,  inner maximization in \eqref{eq:rXb} is solved by ${\bf r}=\frac{{\bf X}{\bf b}}{\left\|{\bf X}{\bf b}\right\|_2}$ and
\begin{equation}
\max_{\left\|{\bf r}\right\|_2=1}{\bf r}^T{\bf X}{\bf b}=\left\|{\bf X}{\bf b}\right\|_2.
\label{eq:Xb2}
\end{equation}
Combining~(\ref{eq:rXb}) and~(\ref{eq:Xb2}), we obtain
\begin{equation}
\max_{\left\|{\bf r}\right\|_2=1}\left\|{\bf X}^T{\bf r}\right\|_1=\max_{{\bf b}\in\{\pm1\}^N}\left\|{\bf X}{\bf b}\right\|_2.
\end{equation}
That is, $
\left\|{\bf X}^T{\bf r}_{L_1}\right\|_1=\left\|{\bf X}{\bf b}_\text{opt}\right\|_2$ 
where ${\bf b}_\text{opt}=\argmax_{{\bf b}\in\{\pm1\}^N}\left\|{\bf X}{\bf b}\right\|_2$ and 
${\bf r}_{L_1}=\frac{{\bf X}{\bf b}_\text{opt}}{\left\|{\bf X}{\bf b}_\text{opt}\right\|_2}$.
\hfill
\IEEEQEDclosed

By Proposition \ref{prop:quad}, to find the principal $L_1$-norm component ${\bf r}_{L_1}$ we solve~(\ref{eq:bopt}) to obtain ${\bf b}_\text{opt}$ and then calculate  $\frac{{\bf X}{\bf b}_\text{opt}}{\left\|{\bf X}{\bf b}_\text{opt}\right\|_2}$.
The straightforward approach to solve (\ref{eq:bopt}) is an exhaustive search among all $2^N$ binary vectors of length $N$.
Therefore, with computational cost $2^N$, Proposition \ref{prop:quad} identifies the $L_1$-optimal principal component of ${\bf X}$.
As the data record size $N$ grows, calculation of the $L_1$ principal component by exhaustive search in (\ref{eq:bopt}) becomes quickly infeasible.
Proposition  \ref{prop:NPhard} below declares that, indeed,  in its general form $\mathcal P_3^{L_1}$, $K=1$, is NP-hard for jointly asymptotically large $N, D$. McCoy and Tropp provide an alternative proof in \cite{McCoy2011},  that is the earliest known to the authors.
\begin{myprop}
The computation of the $L_1$ principal component of ${\bf X}\in{\mathbb R}^{D\times N}$ by maximum $L_1$-norm projection (Problem $\mathcal P_{3}^{L_1}$, $K=1$) is NP-hard in jointly asymptotic $N,D$.
\label{prop:NPhard}
\end{myprop}

\noindent \emph{Proof:}
In~(\ref{eq:rXb}), for any fixed ${\bf r}\in{\mathbb R}^D$, ${\bf b}=\text{sgn}\left({\bf X}^T{\bf r}\right)$.
Hence,
\begin{equation}
{\bf b}_\text{opt}=\text{sgn}\left({\bf X}^T{\bf r}_{L_1}\right).
\label{eq:brL1}
\end{equation}
By~(\ref{eq:rL1}) and~(\ref{eq:brL1}),  computation of the $L_1$ principal component of ${\bf X}$ is equivalent to  computation of ${\bf b}_\text{opt}$ in~(\ref{eq:bopt}).
Consider the special case of~(\ref{eq:bopt}) where ${\bf X}^T{\bf X}={\bf I}_N-{\bf a}{\bf a}^T$, $\mathbf a \in \mathbb R^{N}$, 
$\left\|{\bf a}\right\|_2=1$ (hence, $D=N-1$).
Then,
\begin{equation}
\max_{{\bf b}\in\{\pm1\}^N}{\bf b}^T{\bf X}^T{\bf X}{\bf b}=\max_{{\bf b}\in\{\pm1\}^N}\left\{\left\|{\bf b}\right\|_2^2-\left({\bf b}^T{\bf a}\right)^2\right\}=N-\min_{{\bf b}\in\{\pm1\}^N}\left({\bf b}^T{\bf a}\right)^2.
\end{equation}
But $\min_{\mathbf b \in \{\pm 1 \}^{N}}(\mathbf b^T \mathbf a)^2$ is the NP-complete equal-partition problem~\cite{Garey1979}.
We conclude that  computation of the $L_1$ principal component of ${\bf X}$ is NP-hard in jointly asymptotic $N,D$.
\hfill
\IEEEQEDclosed

\subsection{Existing Approaches in Literature}

Recently there has been a growing documented effort to calculate subspace components by $L_1$ projection maximization~[\citen{Galpin1987}]-[\citen{Gu2012}].
The work in~\cite{Kwak2008} presented a suboptimal iterative algorithm for the computation of ${\bf r}_{L_1}$, which, following the formulation and notation of this present paper, initializes the solution to some arbitrary component $\mathbf r_{L_1}^{(0)}$ and executes
\begin{align}
{\bf b}^{(i+1)}&=\text{sgn}\left({\bf X}^T{\bf r}_{L_1}^{(i)}\right),
\label{eq:Kwak1}\\
{\bf r}_{L_1}^{(i+1)}&=\frac{{\bf X}{\bf b}^{(i+1)}}{\left\|{\bf X}{\bf b}^{(i+1)}\right\|_2},
\label{eq:Kwak2}
\end{align}
$i=0,1,2,\ldots$, until convergence.
The work in~\cite{Nie2011} presented an iterative algorithm for the joint computation of  $K\geq1$ principal $L_1$-norm components.
For the case where $K=1$, the iteration in~\cite{Nie2011} simplifies to the iteration in~\cite{Kwak2008} (that is, (\ref{eq:Kwak1}),~(\ref{eq:Kwak2}) above).
Therefore, for $K=1$, the algorithms in~\cite{Kwak2008}, \cite{Nie2011} are identical and can, in fact, be described by the simple single iteration
\begin{equation}
{\bf b}^{(i+1)}=\text{sgn}\left({\bf X}^T{\bf X}{\bf b}^{(i)}\right),\;\;\;i=1,2,\ldots,
\label{eq:Kwak3}
\end{equation}
for the computation of ${\bf b}_\text{opt}$ in~(\ref{eq:bopt}).
Equation \eqref{eq:Kwak3}, however, does not guarantee convergence to the $L_1$-optimal component solution (convergence to one of the many local maxima may be observed).
In the following section, we present for the first time in the literature an optimal algorithm to calculate the $ L_1$ principal component of a data matrix with complexity polynomial in the sample size $N$ when the data dimension $D$ is fixed.

\subsection{Exact Computation of the $L_1$ Principal Component in Polynomial Time}

Proposition \ref{prop:NPhard} proves  NP-hardness of the computation of the $L_1$ principal component ${\bf r}_{L_1}$ in $N,D$ (that is, when $N,D$ are jointly arbitrarily large).
However, of engineering interest is the case of  fixed data dimension $D$.
In the following, we show for the first time in the literature that, if $D$ is fixed, then  computation of ${\bf r}_{L_1}$ is no longer NP-hard (in $N$).
We state our result in the form of Proposition \ref{prop:polynomial} below.
\begin{myprop}
For any fixed data dimension $D$,  computation of the $L_1$ principal component of ${\bf X}\in{\mathbb R}^{D\times N}$ has complexity ${\mathcal O}\left(N^{\text{{rank}}({\bf X})}\right)$, $\text{{rank}}({\bf X})\leq D$.
\hfill
\IEEEQEDclosed
\label{prop:polynomial}
\end{myprop}

By Proposition \ref{prop:NPhard}, computation of the $ L_1$ principal component of $\mathbf X$ is equivalent to computation of $\mathbf b_{\text{opt}}$ in (\ref{eq:bopt}).
To prove Proposition \ref{prop:polynomial}, we will prove that ${\bf b}_\text{opt}$ can be computed with complexity ${\mathcal O}\left(N^{\text{rank}({\bf X})}\right)$.
We begin our developments by defining
\begin{equation}
d\eqdef\text{rank}({\bf X})\leq D.
\label{eq:d}
\end{equation}
Then, ${\bf X}^T{\bf X}$ also has rank $d$ and can be  decomposed by
\begin{equation}
{\bf X}^T{\bf X}={\bf Q}{\bf Q}^T,\;\;\;{\bf Q}_{N\times d}=\left[{\bf q}_1\;{\bf q}_2\;\ldots\;{\bf q}_d\right],\;\;\;{\bf q}_i^T{\bf q}_j=0,\;i\neq j,
\label{eq:QQ}
\end{equation}
where ${\bf q}_1$, ${\bf q}_2$, $\ldots$ , ${\bf q}_d$ are the $d$ eigenvalue-weighted eigenvectors of ${\bf X}^T{\bf X}$ with nonzero eigenvalue.
By~(\ref{eq:bopt}),
\begin{equation}
{\bf b}_\text{opt}=\argmax_{{\bf b}\in\{\pm1\}^N}{\bf b}^T{\bf Q}{\bf Q}^T{\bf b}=\argmax_{{\bf b}\in\{\pm1\}^N}\left\|{\bf Q}^T{\bf b}\right\|_2.
\label{eq:Qb}
\end{equation}

For the case  $N<D$, the optimal binary vector ${\bf b}_\text{opt}$ can be obtained directly from~(\ref{eq:bopt}) by an exhaustive search among all $2^N$ binary vectors ${\bf b}\in\{\pm1\}^N$.
Therefore, we can design the $L_1$-optimal principal component ${\bf r}_{L_1}$ with computational cost $2^N<2^D={\mathcal O}(1)$.
For the case where the sample size exceeds the data dimension ($N\geq D$), 
we find it useful in terms of both theory and practice to present  our developments separately for data rank $d=1$, $d=2$, and $2 < d \leq D$.\\
\emph{1) Case $d=1$:}
If the data matrix has rank  $d=1$, then ${\bf Q}={\bf q}_1$ and~(\ref{eq:Qb}) becomes
\begin{equation}
{\bf b}_\text{opt}=\argmax_{{\bf b}\in\{\pm1\}^N}\left|{\bf q}_1^T{\bf b}\right|=\text{sgn}\left({\bf q}_1\right).
\end{equation}
By~(\ref{eq:rL1}), the $L_1$-optimal principal component is
\begin{equation}
{\bf r}_{L_1}=\frac{{\bf X}\,\text{sgn}\left({\bf q}_1\right)}{\left\|{\bf X}\,\text{sgn}\left({\bf q}_1\right)\right\|_2}
\label{dequals1}
\end{equation}
designed with complexity ${\mathcal O}\left(N\right)$.
It is of notable practical importance to observe  at this point that even when $\mathbf X$ is not of true rank one, \eqref{dequals1} presents us with a quality, trivially calculated approximation of the $L_1$ principal component of $\mathbf X$:
Calculate the $L_2$ principal component $\mathbf q_1$ of the $N \times N$ matrix $\mathbf X^T \mathbf X$, quantize to $\text{sgn}(\mathbf q_1)$, and project and normalize to obtain $\mathbf r_{L_1} \simeq \mathbf X\,\text{sgn}(\mathbf q_1) / \| \mathbf X\,\text{sgn}(\mathbf q_1)  \|_2$.\\
\emph{2) Case $d=2$:}
If $d=2$, then ${\bf Q}=\left[{\bf q}_1\;\;{\bf q}_2\right]$ and~(\ref{eq:Qb}) becomes
\begin{equation}
{\bf b}_\text{opt}=\argmax_{{\bf b}\in\{\pm1\}^N}\left\{\left({\bf q}_1^T{\bf b}\right)^2+\left({\bf q}_2^T{\bf b}\right)^2\right\}.
\label{eq:rank2}
\end{equation}
The binary optimization problem \eqref{eq:rank2} was seen and solved in \cite{KP} by the auxiliary-angle method \cite{mackenthun},  which was also used earlier in [\citen{sweldens}],[\citen{achilles2}].
Here, we define the $N\times1$ complex vector
\begin{equation}
{\bf z}\eqdef{\bf q}_1+j{\bf q}_2
\end{equation}
and rewrite~(\ref{eq:rank2}) as
\begin{equation}
{\bf b}_\text{opt}=\argmax_{{\bf b}\in\{\pm1\}^N}\left|{\bf b}^T{\bf z}\right|.
\label{eq:bz}
\end{equation}
We introduce the auxiliary angle $\phi\in\left[-\pi,\pi\right)$ and note that, for any complex scalar $w$,
\begin{equation}
\text{Re}\left(we^{-j\phi}\right)\leq|w|
\end{equation}
with equality if and only if $\phi=\text{angle}\left(w\right)$.
That is,
\begin{equation}
|w|=\max_{\phi\in\left[-\pi,\pi\right)}\text{Re}\left(we^{-j\phi}\right).
\end{equation}
Therefore, the maximization in~(\ref{eq:bz}) can be rewritten as
\begin{align}
\max_{{\bf b}\in\{\pm1\}^N}\left|{\bf b}^T{\bf z}\right|&=\max_{{\bf b}\in\{\pm1\}^N}\max_{\phi\in\left[-\pi,\pi\right)}\text{Re}\left({\bf b}^T{\bf z}e^{-j\phi}\right)=\max_{\phi\in\left[-\pi,\pi\right)}\max_{{\bf b}\in\{\pm1\}^N}{\bf b}^T\text{Re}\left({\bf z}e^{-j\phi}\right)
\label{eq:phib}
\end{align}
where, for any given angle $\phi\in\left[-\pi,\pi\right)$,  inner maximization is achieved by
\begin{equation}
{\bf b}\left(\phi\right)=\text{sgn}\left(\text{Re}\left({\bf z}e^{-j\phi}\right)\right).
\label{eq:bphi}
\end{equation}
Then, the optimal vector ${\bf b}_\text{opt}$ in~(\ref{eq:bz}), i.e., the solution to~(\ref{eq:bopt}), is met if we scan the entire interval $\left[-\pi,\pi\right)$ and collect the locally optimal vector ${\bf b}\left(\phi\right)$ for any point $\phi\in\left[-\pi,\pi\right)$.

Interestingly, as we scan the interval $\left[-\pi,\pi\right)$, the locally optimal vector ${\bf b}\left(\phi\right)$ does not change unless the sign of $\text{Re}\left(z_ne^{-j\phi}\right)$ changes for some $n=1,2,\ldots,N$.
Since the latter happens only at $\text{angle}\left(z_n\right)$ and $\text{angle}\left(z_n\right)+\pi$, we obtain  $2N$ points in total at which ${\bf b}\left(\phi\right)$ changes.
Next, we order the $2N$ points with complexity ${\mathcal O}\left(2N\log_22N\right)$ and create successively $2N$ binary vectors by changing each time the sign of $b_n$ if the $n$th element of ${\bf z}$ is the one that determines a sign change.
It is observed that the $2N$ binary vectors that we obtain this way are pair-wise opposite (the vectors that are collected when $\phi\in\left[-\frac{\pi}{2},\frac{\pi}{2}\right)$ are opposite to the ones that are collected when $\phi\in\left[-\pi,-\frac{\pi}{2}\right)\cup\left[\frac{\pi}{2},\pi\right)$).
Since opposite vectors result in the same metric value in~(\ref{eq:bopt}), we can restrict our search to $\left[-\frac{\pi}{2},\frac{\pi}{2}\right)$ and maintain optimality.
Therefore, with overall complexity ${\mathcal O}\left(N\log_2N\right)$, we obtain a set of $N$ binary vectors that contains ${\bf b}_\text{opt}$.
Then, we only have to evaluate the $N$ vectors against the metric of interest in~(\ref{eq:bopt}) to obtain ${\bf b}_\text{opt}$.
We conclude that the $L_1$-optimal principal component of a rank-$2$ matrix ${\bf X}\in{\mathbb R}^{D\times N}$ is designed with complexity ${\mathcal O}\left(N\log_2N\right)$.

\emph{3) Case $d>2$:}
If $d>2$, we design the $L_1$-optimal principal component of ${\bf X}$ with complexity ${\mathcal O}\left(N^d\right)$ by considering the multiple-auxiliary-angle approach that was presented in~\cite{KL} as a generalization of the work in~\cite{KP}.

Consider a unit vector ${\bf c}\in{\mathbbm R}^d$.
By the Cauchy-Schwartz inequality, for any ${\bf a}\in{\mathbbm R}^d$,
\begin{equation}
{\bf a}^T{\bf c}\leq\left\|{\bf a}\right\|_2\left\|{\bf c}\right\|_2=\left\|{\bf a}\right\|_2
\end{equation}
with equality if and only if ${\bf c}$ is codirectional with ${\bf a}$.
Then,
\begin{equation}
\max_{{\bf c}\in{\mathbbm R}^d,\;\left\|{\bf c}\right\|_2=1}{\bf a}^T{\bf c}=\left\|{\bf a}\right\|_2.
\label{eq:ca}
\end{equation}
By~(\ref{eq:ca}), the optimization problem in~(\ref{eq:Qb}) becomes
\begin{equation}
\max_{{\bf b}\in\{\pm1\}^N}\left\|{\bf Q}^T{\bf b}\right\|_2=\max_{{\bf b}\in\{\pm1\}^N}\max_{{\bf c}\in{\mathbbm R}^d,\; \left\|{\bf c}\right\|_2=1}{\bf b}^T{\bf Q}{\bf c}=\max_{{\bf c}\in{\mathbbm R}^d,\; \left\|{\bf c}\right\|_2=1}\max_{{\bf b}\in\{\pm1\}^N}{\bf b}^T{\bf Q}{\bf c}.
\label{eq:maxmax}
\end{equation}
For every ${\bf c}\in{\mathbbm R}^d$,  inner maximization in~(\ref{eq:maxmax}) is solved by the binary vector
\begin{equation}
{\bf b}({\bf c}) = \text{sgn}({\bf Q}{\bf c}),
\label{eq:bc}
\end{equation}
which is obtained with complexity ${\mathcal O}(N)$.
Then, by~(\ref{eq:maxmax}), the solution to the original problem in~(\ref{eq:Qb}) is met if we collect all  binary vectors ${\bf b}({\bf c})$ returned as ${\bf c}$ scans the unit-radius $d$-dimensional hypersphere.
That is, ${\bf b}_\text{opt}$ in~(\ref{eq:Qb}) is in%
\footnote{The $d$th element of vector $\mathbf c$, $c_d$, can be set nonnegative without loss of optimality, because, for any given $\mathbf c$, $\| \mathbf c\|_2=1$, the binary vectors $\mathbf b(\mathbf c)$ and $\mathbf b(\text{sgn}(c_d)\mathbf c)$ result to the same metric value in \eqref{eq:Qb}.}
\begin{equation}
{\mathcal S}_1\eqdef\hspace{-.5cm}\bigcup_{{\bf c}\in{\mathbbm R}^d,\,\left\|{\bf c}\right\|_2=1, \,c_d\geq0}\hspace{-.5cm}{\bf b}({\bf c}).
\label{eq:S1}
\end{equation}
Two fundamental questions for the computational problem under consideration are what the size (cardinality) of set ${\mathcal S}_1$ is and  how much computational effort is expended to form ${\mathcal S}_1$.

We address first the first question. 
We introduce the auxiliary-angle vector ${\boldsymbol\phi} =[\phi_1,\, \phi_2,\, \ldots, \,\phi_{d-1}]^T \in\Phi^{d-1}$,  $\Phi\eqdef\left[-\frac{\pi}{2},\frac{\pi}{2}\right)$, and parametrize ${\bf c}$ as follows, 
\begin{equation}
{\bf c}(\boldsymbol{\phi})\eqdef\begin{bmatrix}
\sin\phi_1\\
\cos\phi_1\sin\phi_2\\
\cos\phi_1\cos\phi_2\sin\phi_3\\
\vdots\\ 
\cos\phi_1\ldots\cos\phi_{d-2}\sin\phi_{d-1}\\
\cos\phi_1\ldots\cos\phi_{d-2}\cos\phi_{d-1}                                                         
\end{bmatrix}.
\end{equation}
Then, we re-express the candidate set in~(\ref{eq:S1}) in the form
\begin{equation}
{\mathcal S}_1=\bigcup_{{\boldsymbol\phi}\in\Phi^{d-1}}{\bf b}({\boldsymbol\phi})
\label{eq:S1phi}
\end{equation}
where, according to~(\ref{eq:bc}),
\begin{equation}
{\bf b}({\boldsymbol\phi}) =
\begin{bmatrix}
b_1({\boldsymbol\phi}), & b_2({\boldsymbol\phi}), & \ldots, & b_{N}({\boldsymbol\phi})
\end{bmatrix}^T
=\text{sgn}({\bf Q}{\bf c}({\boldsymbol\phi})).
\end{equation}
We note that, for any point ${\boldsymbol\phi}$, each element $b_n({\boldsymbol\phi})$, $n=1,2, \ldots, N$, depends only on the corresponding row of ${\bf Q}$ and is determined by $b_n({\boldsymbol{\phi}})=\text{sgn}({\bf Q}_{n,:}\,{\bf c}({\boldsymbol{\phi}}))$.
Hence, the value of the binary element $b_n({\boldsymbol{\phi}})$ changes only when
\begin{equation}
{\bf Q}_{n,:}\,{\bf c}({\boldsymbol{\phi}})=0.
\label{eq:Vnc}
\end{equation}
To gain some insight into the process of introducing the auxiliary-angle vector $\boldsymbol{\phi}$, we notice that the points $\boldsymbol{\phi}$ that satisfy (\ref{eq:Vnc}) determine a hypersurface (or $(d-2)$-manifold) in the $(d-1)$-dimensional space that partitions $\Phi^{d-1}$ into two regions.
One region corresponds to $b_n=-1$ and the other corresponds to $b_n=+1$.
A key observation in the algorithm is that, as ${\boldsymbol{\phi}}$ scans any of the two regions, the decision on $b_n$ does not change.
Therefore, the $N$ rows of ${\bf Q}$ are associated with $N$ corresponding hypersurfaces that partition $\Phi^{d-1}$ into $P_1$ cells $C_1,C_2,\dots,C_{P_1}$ such that $\bigcup_{p=1}^{P_1}C_p=\Phi^{d-1}$, $C_p\cap C_q=\emptyset$ $\forall$ $p\ne q$, and each cell $C_p$ corresponds to a distinct vector ${\bf b}_p\in\{\pm1\}^N$.
As a result, the candidate vector set is $S_1=\bigcup_{p=1}^P\{{\bf b}_p\}$.

\begin{figure*} 
\centering
\includegraphics[width=.8\textwidth]{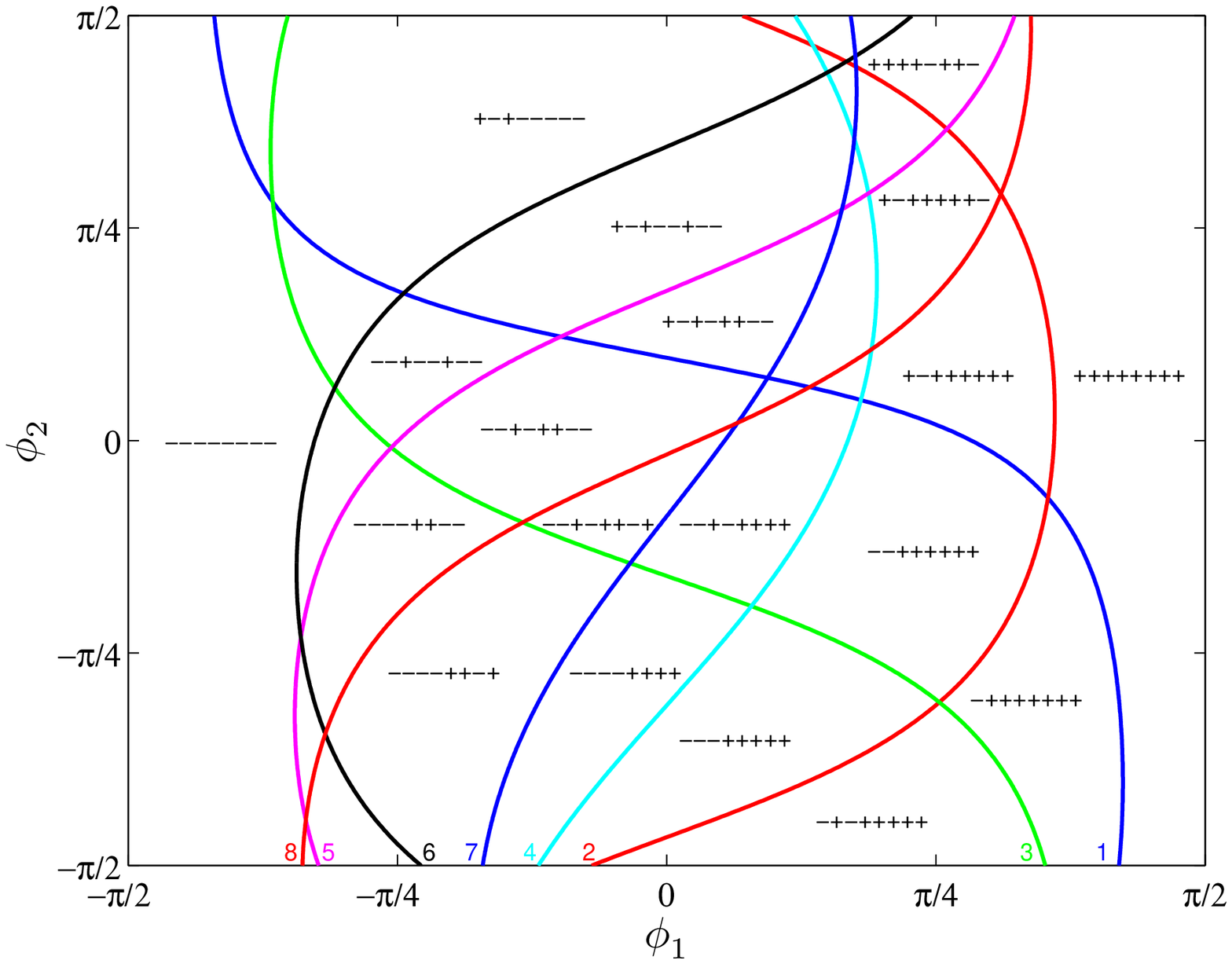}
\caption{Visualization of the calculation of the $L_1$ principal component of a data matrix $\mathbf X_{D \times N}$ of $N=8$ samples with rank $d=3 \leq D$~($D \leq N$).
The space $\Phi^2=\left[ -\frac{\pi}{2}, \frac{\pi}{2} \right) \times \left[ -\frac{\pi}{2}, \frac{\pi}{2} \right)$ is partitioned into $P_1=29$ cells with distinct corresponding binary vectors $\mathbf b_p \in \{ \pm 1 \}^8$, $p=1, 2, \ldots, 29$; $\mathbf b_{\text{opt}}$ in (\ref{eq:bopt}) equals $\mathbf b_p$ for some $p \in \{1, 2, \ldots, 29 \}$ and the $L_1$ principal component is $\mathbf r_{L_1}= \mathbf X \mathbf b_{\text{opt}}/\|\mathbf X \mathbf b_{\text{opt}}\|_2$.}
\label{fig:surfaces}
\end{figure*}

In~\cite{KL}, it was shown that $P_1=\sum_{g=0}^{d-1}\binom{N-1}{g}$ if pairs of cells that correspond to opposite binary vectors (hence, equivalent vectors with respect to the metric of interest in~(\ref{eq:Qb})) are considered as one.
Therefore, the candidate vector set ${\mathcal S}_1$ has cardinality $|{\mathcal S}_1|=\sum_{g=0}^{d-1}\binom{N-1}{g}={\mathcal O}\left(N^{d-1}\right)$.
Fig.~\ref{fig:surfaces} presents a visualization of the algorithm/partition for the case of a data matrix $\mathbf X_{D \times N}$ of $N=8$ samples with rank $d=3\leq D\leq N$.
Since $d=3$, the hypersurfaces (or $(d-2)$-manifolds) are, in fact, curves in the $2$-dimensional space that partition $\Phi^2$ into cells.
The $P=\binom{7}{0} + \binom{7}{1} + \binom{7}{2} = 29 $ cells and associated binary candidate vectors are formed by the eight-row three-column eigenvector matrix ${\bf Q}$ of $\mathbf X^T \mathbf X$ and the scanning angle vector $\mathbf c ({\boldsymbol\phi})=[\sin \phi_1, \; \cos \phi_1 \sin \phi_2, \; \cos \phi_1 \cos \phi_2]^T$.

Regarding the cost of calculating $\mathcal S_1$, since each cell ${ C}$ contains at least one vertex (that is, intersection of $d-1$ hypersurfaces), see for example  Fig.~\ref{fig:surfaces}, it suffices to find all vertices in the partition and determine ${\bf b}$ for all neighboring cells.
Consider $d-1$ arbitrary hypersurfaces; say, for example, ${\bf Q}_{1,:}{\bf c}({\boldsymbol\phi})=0$, ${\bf Q}_{2,:}{\bf c}({\boldsymbol\phi})=0$, $\ldots$, ${\bf Q}_{d-1,:}{\bf c}({\boldsymbol\phi})=0$.
Their intersection satisfies ${\bf Q}_{1,:}{\bf c}({\boldsymbol\phi})={\bf Q}_{2,:}{\bf c}({\boldsymbol\phi})=\ldots={\bf Q}_{d-1,:}{\bf c}({\boldsymbol\phi})=0$ and is computed by solving the equation
\begin{equation}
{\bf Q}_{1:d-1,:}{\bf c}({\boldsymbol\phi})={\bf 0}.
\label{eq:Qc}
\end{equation}
The solution to~(\ref{eq:Qc}) consists of the spherical coordinates of the unit vector in the null space of the $(d-1)\times d$ matrix ${\bf Q}_{1:d-1,:}$.%
\footnote{If ${\bf Q}_{1:d-1,:}$ is full-rank, then its null space has rank $1$ and ${\bf c}({\boldsymbol\phi})$ is uniquely determined (within a sign ambiguity which is resolved by $c_d\geq0$).
If, instead, ${\bf Q}_{1:d-1,:}$ is rank-deficient, then the intersection of the $d-1$ hypersurfaces (i.e., the solution of~(\ref{eq:Qc})) is a $p$-manifold (with $p\geq1$) in the $(d-1)$-dimensional space and does not generate a new cell.
Hence, linearly dependent combinations of $d-1$ rows of ${\bf Q}$ are ignored.}
Then, the binary vector ${\bf b}$ that corresponds to a neighboring cell is computed by 
\begin{equation}
\text{sgn}({\bf Q}\,{\bf c}({\boldsymbol\phi}))
\label{eq:sgnQc}
\end{equation}
with complexity ${\mathcal O}(N)$.
Note that~(\ref{eq:sgnQc}) presents ambiguity regarding the sign of the intersecting $d-1$ hypersurfaces.
A straightforward way to resolve the ambiguity%
\footnote{An alternative way of resolving the sign ambiguities at the intersections of hypersurfaces was developed in~\cite{KL} and led to the direct construction of a set ${\mathcal S}_1$ of size $\sum_{g=0}^{d-1}\binom{N-1}{g}={\mathcal O}(N^{d-1})$ with complexity ${\mathcal O}(N^d)$.}
is to consider all $2^{d-1}$ sign combinations for the corresponding elements $b_1,b_2,\ldots,b_{d-1}$ and obtain the binary vectors of all $2^{d-1}$ neighboring cells.
Finally, we  repeat the above procedure for any combination of $d-1$ intersecting hypersurfaces among the $N$ ones.
Therefore, the total number of binary candidates that we obtain (i.e., the cardinality of ${\mathcal S}_1$) is upper bounded by $2^{d-1}\binom{N}{d-1}={\mathcal O}(N^{d-1})$.
Since complexity ${\mathcal O}(N)$ is required for each combination of $d-1$ rows of ${\bf Q}$ to solve~(\ref{eq:sgnQc}), the overall complexity of the construction of ${\mathcal S}_1$ is ${\mathcal O}(N^d)$ for any given matrix ${\bf Q}_{N\times d}$.

Our complete, new algorithm for the computation of the $L_1$-optimal principal component of a rank-$d$ matrix ${\bf X}\in{\mathbb R}^{D\times N}$ that has complexity ${\mathcal O}\left(N^d\right)$ is presented in detail in Fig.~\ref{fig:algo}.
Computation of each element of $S_1$ (i.e., column of ${\bf B}$ in the algorithm) is performed independently of each other.
Therefore, the proposed algorithm is fully parallelizable.
The space complexity of the algorithm is $O(N)$, since after every computation of a new binary candidate the best binary candidate needs to be stored.

We note that the required optimal binary vector in (\ref{eq:Qb}) can, alternatively, be computed through the algorithm in \cite{Allemand2001}, \cite{Ferrez2005} with time complexity ${\mathcal O}\left(N^{d+1}\right)$ and space complexity at least ${\mathcal O}(N)$ based on the reverse search for cell enumeration in arrangements~\cite{Avis1996} or with time complexity ${\mathcal O}(N^{d-1})$ but space complexity proportional to ${\mathcal O}\left(N^{d-1}\right)$ based on the incremental algorithm for cell enumeration in arrangements \cite{Edelsbrunner1986}, \cite{Edelsbrunner1987}.
Another algorithm that can solve (\ref{eq:Qb}) with polynomial complexity is in \cite{BenAmeur2010}.
Its time complexity is ${\mathcal O}(N^{d-1}\log N)$, while its space complexity is polynomially bounded by the output size (i.e., ${\mathcal O}(N^{d-1})$).
In comparison to the above approaches, the algorithm in Fig. 2 is the fastest known 
with smallest (linear) space complexity.
We conclude that the $L_1$-optimal principal component of a rank-$d$ data matrix ${\bf X}\in{\mathbb R}^{D\times N}$, $d \leq D \leq N$, is obtained with time complexity ${\mathcal O}\left(N^d\right)$ and space complexity ${\mathcal O}(N)$.
That is, the time complexity is polynomial in the sample size with exponent equal to the rank of the data matrix, which is at most equal to the data dimension $D$.
The space complexity is linear in the sample size.

\section{Multiple $L_1$-norm Principal Components}

In this section, we switch our interest to the joint design of  $K>1$ principal $L_1$ components of a $D\times N$ data matrix ${\bf X}$.
After we review suboptimal approaches from the recent literature, we generalize the result of the previous section and prove that, if the data dimension $D$ is fixed, then the $K$ principal $L_1$ components of ${\bf X}$ are computable in polynomial time ${\mathcal O}\left(N^{K \text{rank}(\mathbf X)-K+1}\right)$.

\subsection{Exact Exhaustive-search Computation of Multiple $L_1$ Principal Components}

For any $D\times K$ matrix ${\bf A}$,
\begin{equation}
\max_{{\bf R}\in{\mathbbm R}^{D\times K},\,{\bf R}^T{\bf R}={\bf I}_K}\text{tr}\left({\bf R}^T{\bf A}\right)=\left\|{\bf A}\right\|_*
\label{eq:RAA}
\end{equation}
where $\left\|{\bf A}\right\|_*$ denotes the nuclear norm (i.e., the sum of the singular values) of ${\bf A}$.
Maximization in~(\ref{eq:RAA}) is achieved by ${\bf R}={\bf U}{\bf V}^T$ where ${\bf U}{\bf\Sigma}{\bf V}^T$ is the ``compact'' SVD of ${\bf A}$, ${\bf U}$ and ${\bf V}$ are $D\times d$ and $K\times d$, respectively, matrices with ${\bf U}^T{\bf U}={\bf V}^T{\bf V}={\bf I}_d$, ${\bf\Sigma}$ is a nonsingular diagonal $d\times d$ matrix, and $d$ is the rank of ${\bf A}$.
This is due to the trace version of the Cauchy-Schwarz inequality~\cite{Cauchy} according to which
\begin{align}
\text{tr}\left({\bf R}^T{\bf A}\right)&=\text{tr}\left({\bf R}^T{\bf U}{\bf\Sigma}{\bf V}^T\right)=\text{tr}\left({\bf U}{\bf\Sigma}^\frac{1}{2}\cdot{\bf\Sigma}^\frac{1}{2}{\bf V}^T{\bf R}^T\right) \nonumber \\
&\leq\left\|{\bf U}{\bf\Sigma}^\frac{1}{2}\right\|_2\left\|{\bf\Sigma}^\frac{1}{2}{\bf V}^T{\bf R}^T\right\|_2=\left\|{\bf\Sigma}^\frac{1}{2}\right\|_2^2=\text{tr}\left({\bf\Sigma}\right)=\left\|{\bf A}\right\|_*
\label{eq:SA}
\end{align}
with equality if $\left({\bf U}{\bf\Sigma}^\frac{1}{2}\right)^T={\bf\Sigma}^\frac{1}{2}{\bf V}^T{\bf R}^T$ which is satisfied by ${\bf R}={\bf U}{\bf V}^T$.

To identify the optimal $L_1$ subspace for any number of  components $K$, we begin by presenting a property of ${\mathcal P}^{L_1}_3$ in the form of Proposition~\ref{prop:nuclear} below. Proposition \ref{prop:nuclear} is  a generalization of Proposition~\ref{prop:quad} and interprets ${\mathcal P}^{L_1}_3$ as an equivalent nuclear-norm maximization problem over the binary field.
\begin{myprop}
For any data matrix $\mathbf X \in \mathbb R^{D \times N}$, the solution to 
${\mathcal P}^{L_1}_3:\;\;\;\;{\bf R}_{L_1}=\argmax_{{\bf R}\in{\mathbb R}^{D\times K}, \; {\bf R}^T{\bf R}={\bf I}_K}\left\|{\bf R}^T{\bf X}\right\|_1$
is given by
\begin{equation}
{\bf R}_{L_1}={\bf U}{\bf V}^T
\end{equation}
where ${\bf U}$ and ${\bf V}$ are the $D\times K$ and $N\times K$ matrices that consist of the $K$ dominant-singular-value left and right, respectively, singular vectors of ${\bf X}{\bf B}_\text{{opt}}$ with
\begin{equation}
{\bf B}_\text{{opt}}=\argmax_{{\bf B}\in\{\pm1\}^{N\times K}}\left\|{\bf X}{\bf B}\right\|_*.
\label{eq:Bopt}
\end{equation}
In addition,
$\left\|{\bf R}_{L_1}^T{\bf X}\right\|_1=\left\|{\bf X}{\bf B}_\text{{opt}}\right\|_*$.
\label{prop:nuclear}
\end{myprop}

\noindent \emph{Proof:} We  rewrite the optimization problem in~(\ref{eq:RL1}) as
\begin{align}
\max_{\mathbf R \in \mathbb R^{D \times K}, \; {\bf R}^T{\bf R}={\bf I}_K}\left\|{\bf X}^T{\bf R}\right\|_1 &=\max_{\mathbf R \in \mathbb R^{D \times K}, \;  {\bf R}^T{\bf R}={\bf I}_K}\sum_{k=1}^K\left\|{\bf X}^T{\bf r}_k\right\|_1=\max_{\mathbf R \in \mathbb R^{D \times K}, \; {\bf R}^T{\bf R}={\bf I}_K}\sum_{k=1}^K\max_{{\bf b}_k\in\{\pm1\}^N}{\bf b}_k^T{\bf X}^T{\bf r}_k \nonumber \\
&=\max_{\mathbf R \in \mathbb R^{D \times K}, \; {\bf R}^T{\bf R}={\bf I}_K}\max_{{\bf B}\in\{\pm1\}^{N\times K}}\text{tr}\left({\bf B}^T{\bf X}^T{\bf R}\right) \nonumber \\
 &=\max_{{\bf B}\in\{\pm1\}^{N\times K}}\max_{\mathbf R \in \mathbb R^{D \times K}, \; {\bf R}^T{\bf R}={\bf I}_K}\text{tr}\left({\bf R}^T{\bf X}{\bf B}\right) = \max_{{\bf B}\in\{\pm1\}^{N\times K}}\left\|{\bf X}{\bf B}\right\|_*. 
 \label{eq:RXB} 
\end{align}
That is, $\left\|{\bf R}_{L_1}^T{\bf X}\right\|_1=\left\|{\bf X}{\bf B}_\text{{opt}}\right\|_*$ where ${\bf B}_\text{opt}=\argmax_{{\bf B}\in\{\pm1\}^{N\times K}}\left\|{\bf X}{\bf B}\right\|_*$ and, by~(\ref{eq:RAA}) and~(\ref{eq:SA}), ${\bf R}_{L_1}={\bf U}{\bf V}^T$ where 
${\bf U}{\bf\Sigma}{\bf V}^T$ is the ``compact'' SVD of ${\bf X}{\bf B}_\text{opt}$.
\hfill
\IEEEQEDclosed

By Proposition~\ref{prop:nuclear}, to find exactly the optimal $L_1$-norm projection operator ${\bf R}_{L_1}$ we can perform the following steps:
\begin{enumerate}
\item
Solve~(\ref{eq:Bopt}) to obtain ${\bf B}_\text{opt}$.
\item
Perform SVD on ${\bf X}{\bf B}_\text{opt}={\bf U}{\bf\Sigma}{\bf V}^T$.
\item
Return ${\bf R}_{L_1}={\bf U}_{:,1:K}{\bf V}^T$.
\end{enumerate}
Steps $1$ - $3$ offer for the first time a direct approach for the computation of the $K$ jointly-optimal $L_1$ principal components of ${\bf X}$.
Step $1$ can be executed by an exhaustive search among all $2^{NK}$ binary matrices of size $N\times K$ followed by evaluation in  the metric of interest in~(\ref{eq:Bopt}).
That is, with computational cost $\mathcal O(2^{NK})$ we identify the $L_1$-optimal $K$ principal components of ${\bf X}$.

\subsection{Existing Approaches in Literature}

For the case $K>1$,~\cite{Kwak2008} proposed to design the first $L_1$ principal component ${\bf r}_{L_1}$ by the coupled  iteration~(\ref{eq:Kwak1})-(\ref{eq:Kwak2}) (which does not guarantee optimality) and then project the data onto the subspace that is orthogonal to ${\bf r}_{L_1}$; design the $L_1$ principal component of the projected data by the same coupled iteration; and continue similarly.
To avoid the above suboptimal projection-greedy approach,~\cite{Nie2011} presented an iterative algorithm for the computation of ${\bf R}_{L_1}$ altogether (that is the joint computation of the $K$ principal $L_1$ components). In the language of Proposition \ref{prop:nuclear},
the algorithm can be described as arbitrary
initialization at some ${\bf R}_{L_1}^{(0)}$ followed by updates
\begin{align}
{\bf B}^{(i+1)}&=\text{sgn}\left({\bf X}^T{\bf R}_{L_1}^{(i)}\right),\label{eq:UTA1}\\
\left({\bf U}^{(i+1)},{\bf\Sigma}^{(i+1)},{\bf V}^{(i+1)}\right)&=\text{SVD}\left({\bf X}{\bf B}^{(i+1)}\right),\\
{\bf R}_{L_1}^{(i+1)}&={\bf U}_{:,1:K}^{(i+1)}{{\bf V}^{(i+1)}}^T\label{eq:UTA3},
\end{align}
for $i=0, 1,2,\ldots$, until convergence.
Similar to the work in~\cite{Kwak2008}, the above iteration does not guarantee convergence to the $L_1$-optimal subspace.

\subsection{Exact Computation of Multiple $L_1$ Principal Components in Polynomial Time}

By the proof of Proposition~\ref{prop:nuclear},  for any given ${\bf R}\in{\mathbb R}^{D\times K}$ the corresponding metric-maximizing binary matrix is ${\bf B}=\text{sgn}\left({\bf X}^T{\bf R}\right)$. Hence,
\begin{equation}
{\bf B}_\text{opt}=\text{sgn}\left({\bf X}^T{\bf R}_{L_1}\right).
\label{eq:BRL1}
\end{equation}
By Proposition~\ref{prop:nuclear} and~(\ref{eq:BRL1}),   computation of the $K$ principal $L_1$ components of ${\bf X}_{D \times N}$ is equivalent to  computation of ${\bf B}_\text{opt}$ in~(\ref{eq:Bopt}), which indicates NP-hardness in $N,D$ (that is, when $N,D$ are arbitrarily large). As before,
in this section we consider the case of engineering interest of fixed data dimension $D$.
As in Section~\ref{sec:OneComponent}, we show that, if $D$ is fixed, then computation of the $K$ principal $L_1$ components of ${\bf X}$ is no longer NP-hard (in $N$).
We state our result in the form of the following proposition.

\begin{myprop}
For any fixed data dimension $D$, optimal computation of the $K$ principal $L_1$ components of ${\bf X}\in{\mathbb R}^{D\times N}$ can be carried out with complexity ${\mathcal O}\left(N^{\text{{rank}}({\bf X})K-K+1}\right)$, $\text{{rank}}({\bf X})\leq D$.
\hfill
\IEEEQEDclosed
\label{prop:polynomial_multiple}
\end{myprop}
To prove Proposition~\ref{prop:polynomial_multiple}, it suffices to prove that ${\bf B}_\text{opt}$ can be computed with complexity ${\mathcal O}\left(N^{\text{{rank}}({\bf X})K-K+1}\right)$.
As in~(\ref{eq:d}), (\ref{eq:QQ}), let $d$ denote the rank of ${\bf X}$ and ${\bf Q}{\bf Q}^T$ where ${\bf Q}\in{\mathbbm R}^{N\times d}$ is the eigen-decomposition matrix of ${\bf X}^T{\bf X}$.
By~(\ref{eq:Bopt}),
\begin{equation}
{\bf B}_\text{{opt}}=\argmax_{{\bf B}\in\{\pm1\}^{N\times K}}\sum_{k=1}^K\sqrt{\lambda_k\left[{\bf B}^T{\bf X}^T{\bf X}{\bf B}\right]}=\argmax_{{\bf B}\in\{\pm1\}^{N\times K}}\sum_{k=1}^K\sqrt{\lambda_k\left[{\bf B}^T{\bf Q}{\bf Q}^T{\bf B}\right]}=\argmax_{{\bf B}\in\{\pm1\}^{N\times K}}\left\|{\bf Q}^T{\bf B}\right\|_*
\label{eq:QB}
\end{equation}
where  $\lambda_k [{\bf A}]$ denotes the $k$th  eigenvalue of matrix ${\bf A}$, $k=1, \ldots, K$.

For the case $N<D$, the optimal binary matirx ${\bf B}_\text{opt}$ can be obtained directly from~(\ref{eq:Bopt}) by an exhaustive search among all $2^{NK}$ binary matrices ${\bf B}\in\{\pm1\}^{N\times K}$.
Therefore, we can design the $L_1$-optimal $K$ principal components with computational cost $2^{NK}<2^{DK}={\mathcal O}(1)$.

For the (certainly more interesting) case where the sample size exceeds the data dimension, $N\geq D$, we present for the first time a generalized version of the approach in~\cite{KP},~\cite{KL} that introduces an orthonormal scanning matrix  to maximize a rank-deficient nuclear norm.
In particular, we observe by~(\ref{eq:QB}) that we need ${\bf B}_\text{{opt}}$ that solves
\begin{equation}
\max_{{\bf B}\in\{\pm1\}^{N\times K}}\left\|{\bf Q}^T{\bf B}\right\|_*\overset{(\ref{eq:RAA})}{=} \max_{{\bf B}\in\{\pm1\}^{N\times K}}\max_{{\bf C}\in{\mathbbm R}^{d\times K},\,{\bf C}^T{\bf C}={\bf I}_K}\text{tr}\left({\bf C}^T{\bf Q}^T{\bf B}\right)=\max_{{\bf C}\in{\mathbbm R}^{d\times K},\,{\bf C}^T{\bf C}={\bf I}_K}\max_{{\bf B}\in\{\pm1\}^{N\times K}}\text{tr}\left({\bf B}^T{\bf Q}{\bf C}\right).
\label{eq:BQC}
\end{equation}
By interchanging the maximizations  in~(\ref{eq:BQC}), for any fixed $d\times K$ matrix ${\bf C}$ the inner maximization with respect to $\mathbf B \in \{ \pm 1 \}^{N \times K}$ is solved by
\begin{equation}
{\bf B}({\bf C}) = \left[\text{sgn}({\bf Q}{\bf C}_{:,1}), \;\text{sgn}({\bf Q}{\bf C}_{:,2}), \;\ldots, \;\text{sgn}({\bf Q}{\bf C}_{:,K})\right],
\label{eq:BC}
\end{equation}
which is obtained with complexity linear in $N$.
Then, by~(\ref{eq:BQC}), the solution to our original problem in~(\ref{eq:QB}) is met if we  collect all possible binary matrices ${\bf B}({\bf C})$ returned as the columns of ${\bf C}$ scan the unit-radius $d$-dimensional hypersphere while maintaining orthogonality among them.
That is, ${\bf B}_\text{opt}$ in~(\ref{eq:QB}) is in%
\footnote{Without loss of optimality, we set $C_{d,k} \geq 0$,  $k =1, 2, \ldots, K$, since, for any given $\mathbf C$, $\mathbf C^T \mathbf C = \mathbf I_K$, the binary matrices $\mathbf B(\mathbf C)$ and $\mathbf B\big( \mathbf C \, \text{diag} ( \text{sgn}(\mathbf C_{d, :})  ) \big) $ result to the same metric value in \eqref{eq:Bopt}.}
\begin{equation}
{\mathcal S}_K\eqdef\hspace{-1cm}\bigcup_{\begin{smallmatrix}{\bf C}\in{\mathbbm R}^{d\times K},\,{\bf C}^T{\bf C}={\bf I}_K,\\C_{d,k}\geq0,\,k=1,2,\ldots,K\end{smallmatrix}}\hspace{-1cm}{\bf B}({\bf C}).
\end{equation}
Then, by relaxing  orthogonality among the columns of ${\bf C}$, 
\begin{equation}
{\mathcal S}_K\subset\hspace{-1cm}\bigcup_{\begin{smallmatrix}{\bf C}\in{\mathbbm R}^{d\times K},\,\left[{\bf C}^T{\bf C}\right]_{k,k}=1,\\C_{d,k}\geq0,\,k=1,2,\ldots,K\end{smallmatrix}}\hspace{-1cm}{\bf B}({\bf C}){=}\Big(\bigcup_{\begin{smallmatrix}{\bf c}\in{\mathbbm R}^d,\,\left\|{\bf c}\right\|_2=1,\\c_d\geq0\end{smallmatrix}}\hspace{-.5cm}{\bf b}({\bf c})\Big)^K{=}{\mathcal S}_1\times{\mathcal S}_1\times\ldots\times{\mathcal S}_1={\mathcal S}_1^K,
\label{eq:S1K}
\end{equation}
which implies that
\begin{equation}
\left|{\mathcal S}_K\right|\leq\left|{\mathcal S}_1\right|^K=\left({\mathcal O}\left(N^{d-1}\right)\right)^K={\mathcal O}\left(N^{dK-K}\right).
\label{eq:SKO}
\end{equation}
From~(\ref{eq:SKO}), we observe that the number of binary matrices that we collect as the columns of ${\bf C}$ scan the unit-radius $d$-dimensional hypersphere --with or without maintaining orthogonality-- is polynomial in $N$.
After ${\bf C}$ has finished scanning the hypersphere, all collected binary matrices in ${\mathcal S}_K$ are compared to each other against the metric of interest in~(\ref{eq:QB}) with complexity ${\mathcal O}(N)$ per matrix.
Therefore, the complexity to solve~(\ref{eq:Bopt}) is determined by the complexity to build ${\mathcal S}_K$ or at most $\mathcal S_1^K$ since $\mathcal S_K \subset \mathcal S_1^K$ by \eqref{eq:S1K}.

Since ${\bf B}_\text{opt}\in{\mathcal S}_K{\subset}{\mathcal S}_1^K$, we already have a direct way to solve~(\ref{eq:QB}).
First, we construct ${\mathcal S}_1$ with complexity ${\mathcal O}\left(N^d\right)$ as described in Section~\ref{sec:OneComponent}.
We note that ${\mathcal S}_1$ contains ${\mathcal O}\left(N^{d-1}\right)$ binary vectors.
Then, we  construct ${\mathcal S}_1^K$ which consists of all selections of $K$ elements of ${\mathcal S}_1$ allowing repeated elements.
The order of the elements in each selection can be disregarded, since the order of the columns of ${\bf B}$ does not affect the metric in~(\ref{eq:QB}).
Hence, the total number of selections that we need to consider is the number of possible ways one can choose $K$ elements from a set of $\left|{\mathcal S}_1\right|$ elements disregarding order and allowing repetitions (i.e., the number of size-$K$ multisets of all ${\mathcal S}_1$), which equals~\cite{stanley}
\begin{equation}
P_K=\binom{\left|{\mathcal S}_1\right|+K-1}{K}={\mathcal O}\left(N^{dK-K}\right)
\end{equation}
since $|{\mathcal S}_1|={\mathcal O}\left(N^{d-1}\right)$.
For each one of the $P_K$ binary matrices, we evaluate the corresponding metric $\left\|{\bf Q}^T{\bf B}\right\|_*$ in~(\ref{eq:QB}) with complexity ${\mathcal O}(N)$.
Then, we identify the optimal matrix ${\bf B}_\text{opt}$ by comparing the calculated metric values.
Therefore, the overall complexity to solve~(\ref{eq:Bopt}) is ${\mathcal O}\left(N^{Kd-K}\right)\cdot{\mathcal O}\left(N\right)={\mathcal O}\left(N^{dK-K+1}\right)$.

The complete  algorithm for the computation of the optimal $K$-dimensional ($K>1$) $L_1$-principal subspace of a rank-$d$ matrix ${\bf X}\in{\mathbb R}^{D\times N}$ with complexity ${\mathcal O}\left(N^{dK - K +1 }\right)$ is given in Fig.~\ref{fig:algo_multi}. 
As a simple illustration of the practical computational cost of the presented algorithm, in Table  I  we show the average CPU time expended by an Intel$^{{\tiny \textregistered}}$ Core$^{{\tiny \texttrademark}}$ i5 Processor at 3.40 GHz running the algorithm of Fig. \ref{fig:algo_multi} in Matlab$^{{\tiny \textregistered}}$ R2012a  to calculate  the $K=2$ principal components of a $d \times N$ rank-$d$ data matrix for $d=3,4, 5, 6$ and $N = 4,6, \ldots, 14$ (we consider only the cases $N>d$).
The presented CPU time for each $(d, N)$ case is the average over $100$ data matrix realizations
created with independent zero-mean unit-variance Gaussian drawn entries.
Importantly, per Figs. 2 and 3, both visiting the ${N \choose d-1}$ manifold-intersection points  for constructing $\mathcal S_1$  (lines 2-8 of function \emph{compute\_candidates} in Fig. \ref{fig:algo}) and  constructing  $\mathcal S_K$ given $\mathcal S_1$ (line 4 of the $L_1$-principal subspace algorithm in Fig. \ref{fig:algo_multi}) are fully parallelizable actions that can be distributed over multiple processing units.
Thus, the entire subspace calculation is  fully parallelizable and the expended calculation time can be divided down by the number of available processors (plus necessary inter-processor communication overhead).

\section{Experimental Studies}

In this section, we carry out a few experimental studies on  $L_1$-subspace signal processing to motivate and illustrate the theoretical developments in the previous sections.
Examples are drawn from the research fields of dimensionality reduction, data restoration,  direction-of-arrival estimation, and image conditioning/reconstruction.

\subsection*{Experiment 1 - Data Dimensionality Reduction}

We generate a nominal data set $\mathbf X_{D \times N}$ of $N=50$  two-dimensional ($D=2$)  observation points  drawn from the Gaussian distribution $\mathcal N \left( \mathbf 0_2, 
\begin{bmatrix}
15 & 13 \\ 13 & 26
\end{bmatrix}
\right)$  as seen in Fig. \ref{fig:dr1}.
We calculate and plot in Fig. \ref{fig:dr1} the $L_2$ (by standard SVD) and $L_1$ (by Section III.C, Case $d=2$, complexity about $50 \log_2 50$) principal component of the data matrix $\mathbf X$.%
\footnote{We note that without the presented algorithm, computation of the $L_1$ principal component of $\mathbf X_{2 \times 50}$ would have required complexity proportional to $2^{50}$ (by \eqref{eq:rank2}), which is of course infeasible.}
For reference purposes, we also plot the true nominal data maximum-variance direction, i.e., the 
dominant eigenvector of the autocorrelation matrix 
$\begin{bmatrix}
15 & 13 \\ 13 & 26
\end{bmatrix}$.
Then, we assume that our data matrix is corrupted by four outlier measurements, $\mathbf o_1, \mathbf o_2, \mathbf o_3, \mathbf o_{4}$, shown in the bottom right corner of Fig. \ref{fig:dr2}.
We recalculate the $L_2$ and $L_1$ principal component of the corrupted data matrix $\mathbf X^{\text{CRPT}} = [\mathbf X, \mathbf o_1, \mathbf o_2 , \mathbf o_3, \mathbf o_{4}]$ and notice (Fig. \ref{fig:dr1} versus Fig. \ref{fig:dr2}) how strongly the $L_2$ component responds to the outliers compared to $L_1$.
To quantify the impact of the outliers, in Fig. \ref{fig:dr3} we generate $1000$ new independent evaluation data points from $\mathcal N \left( \mathbf 0_2, 
\begin{bmatrix}
15 & 13 \\ 13 & 26
\end{bmatrix}
\right)$
and estimate the mean square-fit-error  $\text{E} \left\{ \|\mathbf x - \mathbf r\mathbf r^T \mathbf x \|_2^2 \right\}$ when $\mathbf r=\mathbf r_{L_2}(\mathbf X^{\text{CRPT}})$ or $\mathbf r_{L_1}(\mathbf X^{\text{CRPT}})$.
We find $\frac{1}{1000} \sum_{i=1}^{1000} \| {\mathbf x}_i  - \mathbf r_{L_2}(\mathbf X^{\text{CRPT}})\mathbf r_{L_2}(\mathbf X^{\text{CRPT}})^T {\mathbf x}_i \|_2^2  = 34.417$ versus  $\frac{1}{1000} \sum_{i=1}^{1000} \| {\mathbf x}_i  - \mathbf r_{L_1}(\mathbf X^{\text{CRPT}})\mathbf r_{L_1}(\mathbf X^{\text{CRPT}})^T {\mathbf x}_i \|_2^2 = 11.555$.
In contrast, when the principal component is calculated from the clean training set, $\mathbf r= \mathbf r_{L_2}(\mathbf X)$ or $ \mathbf r_{L_1}(\mathbf X)$, we find estimated mean square-fit-error $6.077$ and $ 6.080$, correspondingly.
We conclude that dimensionality reduction by $L_1$ principal components may loose only minimally in mean-square fit compared to $L_2$ when the designs are from clean training sets, but can protect significantly when training is carried out in the presence of erroneous data. 

Next, we will compare the dimensionality-reduction performance of the proposed $L_1$-principal subspace with that of other subspaces in the literature obtained by means of $L_1$-norm based methods.
Specifically, alongside the $L_2$ (SVD) and $L_1$-principal component (proposed),
we calculate the $R_1$-principal component \cite{Ding2006} as well as the direction obtained by means of $L_1$-factorization through alternating weighted median calculation \cite{Ke2003}, \cite{Ke2005}.\footnote{Notice that for $R_1$-PCA \cite{Ding2006} and $L_1$-factorization \cite{Ke2003}, \cite{Ke2005}, no optimal solution exists in the literature so far.} 
All directions are calculated from an $(N=20)$-point corrupted data set $\mathbf X^{\text{CRPT}} \in \mathbb R^{2 \times 20}$ with  $N_{\mathrm{out}}$ outliers drawn from $\mathcal N \left( \begin{bmatrix} 20 \\ -20  \end{bmatrix},   \begin{bmatrix}    5.73   & -4.494  \\   -4.494  &   5.27   \end{bmatrix} \right)$ and $N - N_{\mathrm{out}}$ nominal points drawn from 
$\mathcal N \left( \mathbf 0_{3}, \begin{bmatrix} 15 & 13 \\ 13 & 26 \end{bmatrix} \right)$.
In Fig. \ref{fig:linefitting_mse}, we plot the mean-squared-fit-error 
averaged over $10 000$ independent corrupted training data-set experiments
as a function of the number of outlying points in the data set $N_{\mathrm{out}}$. We notice that, when  designed on nominal  data, all examined subspaces  differ little, if any,  from the $L_2$-principal subspace   in mean-square  fit error.
However,  when  designed on outlier-corrupted data sets, the  $L_1$-principal subspace exhibits  
notable robustness outperforming uniformly and significantly all other subspaces, especially
in the $15\%$ - $40\%$ mid-range of corruption. Given that $L_1$ and $L_2$ start very near 
each other
in mean-square-fit-error at $0\%$ corruption and meet again only at $100\%$ corruption, 
one is tempted
to say that the $L_1$ subspaces are to be uniformly preferred over $L_2$ if the associated
computational cost can be afforded.

\renewcommand{\arraystretch}{0.6}
\subsection*{Experiment 2 - Data Restoration}
As a toy numerical example, consider a hypothetical case where we collect from a sensor system
eight samples of five-dimensional data. Due to the nature of the sensed source, the data are to 
lie in a lower-than-five dimensional space, say a plane. Say, then, the true data are given by the
$\text{rank}$-$2$ data matrix below 
{\footnotesize
\begin{align}
\mathbf X_{5 \times 8} = 
\begin{bmatrix}
    2.0724  &  -1.2024 &   1.2956 &   2.8719 &   1.5637 &  -2.9323 &  -3.1792 &  -1.4152 \\
   -0.5233 &   0.2595  & -0.3298  & -0.7562 &  -0.4087  &  0.7973  &  0.8235  &  0.4155 \\
    0.0185 &  -0.8158  &  -0.0367 &  -0.5406 &  -0.2380 &   1.0108 &   0.3502 &   1.0487 \\
   -0.6424  &  0.1476  & -0.4151  & -1.0486 &  -0.5552  &  1.1989  &  1.0913  &  0.7355 \\
   -2.1289  &  2.2734  & -1.2687 &  -2.2200 &  -1.2814  &  1.6751 &   2.7777  &  0.0851 \\
 \end{bmatrix}. \nonumber
\end{align}
}
Assume that due to sensor malfunction or data transfer error or data storage failure, 
we are presented instead with 
{\footnotesize
\begin{align}
\mathbf X_{5 \times 8}^{\text{CRPT}} = 
\begin{bmatrix}
    2.0724   & {\color{red}\it{ 8.9538}}  &  1.2956 &   2.8719  & {\color{red}\it{ 10.6817}} &   -2.9323  & -3.1792 &   -1.4152 \\
   -0.5233 & {\color{red}\it{  10.6187}}  & -0.3298 &  -0.7562 &  {\color{red}\it{ 11.0235}} &   0.7973  &  0.8235 &   0.4155 \\
    0.0185 & {\color{red}\it{  11.3050}}  & -0.0367 &  -0.5406 &  -0.2380 &   1.0108  &  0.3502  &  1.0487 \\
   -0.6424  &  0.1476  & -0.4151 &  -1.0486  & {\color{red}\it{  7.8846}}  &  1.1989  &  1.0913  &  0.7355 \\
   -2.1289  &  2.2734 &  -1.2687 &  -2.2200 &  -1.2814  &  1.6751 &   2.7777  &  0.0851 \\
 \end{bmatrix}\nonumber
\end{align}
}\\
where six of the original entries in two of the data points have been 
altered/overwritten and  $\mathbf X^{\text{CRPT}} $ spans now a four-dimensional subspace of $\mathbb R^{5}$.

Our objective is to ``restore'' $\mathbf X^{\text{CRPT}} $ to $\mathbf X$ taking advantage of
our knowledge (or assumption) of the rank of the original data. Along these lines,
we project  $\mathbf X^{\text{CRPT}}$ onto the span of its $K=2$ $L_2$- or $L_1$-principal components,
\begin{align}
\hat{\mathbf X}=\mathbf R \mathbf R^T \mathbf X^{\text {CRPT}}
\end{align}
where $\mathbf R_{5 \times 2}= [\mathbf r_{L_2}^{(1)}, \mathbf r_{L_2}^{(2)}]$ or $ [\mathbf r_{L_1}^{(1)}, \mathbf r_{L_1}^{(2)}]$.
The resulting $L_2$- and $L_1$-derived representations of  $\mathbf X$ are
{\footnotesize
\begin{align}
\hat{\mathbf X}_{L_2}= 
\begin{bmatrix}
    0.8029 &   8.2311 &   0.4919 &   0.9945 &  11.8445 &  -0.9197 &  -1.1528 &  -0.3268\\
    0.4839 &  11.0891 &   0.2888 &   0.5096 &  10.2500  & -0.3897 &  -0.6347 &  -0.0285\\
   -0.5922 &  11.1679 &  -0.3843 &  -0.9862 &   0.0165  &  1.1412 &   1.0192  &  0.7148\\
    0.6521 &   0.8969 &   0.4067 &   0.8926 &   6.6810 &   -0.9024 &  -0.9930 &  -0.4245\\
   -0.3868 &  2.8347 &  -0.2455 &  -0.5789 &  -2.2540  &  0.6257  &  0.6220  &  0.3444
 \end{bmatrix} \nonumber
\end{align}
}
and
{\footnotesize
\begin{align}
\hat{\mathbf X}_{L_1}= 
\begin{bmatrix}
    2.0724 &  -0.0303 &   1.2956 &   2.8719  &  2.9321 &  -2.9323  & -3.1792  & -1.4152\\
   -0.5233 &   0.1880 &  -0.3298 &  -0.7562  & -0.7283 &   0.7973 &   0.8235  &  0.4155\\
    0.0185 &   3.2915 &  -0.0367 &  -0.5406 &   0.2476 &   1.0108 &   0.3502 &   1.0487\\
   -0.6424 &   0.9300 &  -0.4151 &  -1.0486 & -0.8469  &  1.1989  &  1.0913   & 0.7355\\
   -2.1289 &  -4.2139 &  -1.2687 &  -2.2200 &  -3.2976  &  1.6751 &   2.7777 &   0.0851
 \end{bmatrix}, \nonumber
\end{align}
}

\noindent{respectively}.
In Fig.~\ref{fig:error}, we plot the element-by-element and per-measurement  square-restoration error for the two projections.
The relative superiority of $L_1$-subspace data representation is clearly captured and documented.
        
\subsection*{Experiment 3 - Direction-of-Arrival Estimation} 

We consider a uniform linear antenna array of $D=5$ elements that takes $N=10$ snapshots of two incoming signals with angles of arrival  $\theta_1 = -30^\circ$ and $\theta_2 = 50^\circ$, 
\begin{align}
\mathbf x_n= A_1  \mathbf s_{\theta_1} + A_2   \mathbf s_{\theta_2} + \mathbf n_n, ~n=1, \ldots, 10,  \label{cleanmes}
\end{align}
where $A_1, A_2$ are the received-signal amplitudes with array response vectors $\mathbf s_{\theta_1}$ and $\mathbf s_{\theta_2}$, correspondingly, and 
$\mathbf n \sim \mathcal {CN}\left(\mathbf 0_{5}, \sigma^2 \mathbf I_{5} \right)$ is  additive white complex Gaussian noise.
We assume that the signal-to-noise ratio (SNR) of the two signals is  $\text{SNR}_1=10 \log_{10}\frac{A_1^2}{\sigma^2}\text{dB}=2\text{dB}$ and $\text{SNR}_2=10 \log_{10}\frac{A_2^2}{\sigma^2}\text{dB}=3\text{dB}$.
Next, we assume that one arbitrarily selected measurement  out of the ten  observations $\mathbf X_{5 \times 10}=[\mathbf x_1, \ldots, \mathbf x_{10}] \in \mathbb C^{5 \times 10}$ is corrupted by a jammer operating at angle  $\theta_J= 20^\circ$ with amplitude $A_J = A_2$.
We call the resulting corrupted observation set  $\mathbf X^{\text{CRPT}} \in \mathbb C^{5 \times 10}$ and create the real-valued version  $\tilde{\mathbf X}^{\text{CRPT}} = [ \text{Re} \{\mathbf X^{\text{CRPT}}\}^T, ~ \text{Im} \{\mathbf X^{\text{CRPT}}\}^T ]^T \in \mathbb R^{10 \times 10}$ by $\text{Re}\{ \cdot\}, \text{Im}\{ \cdot\} $ part concatenation.
We calculate the $K=2$ $L_2$-principal components of $\tilde{\mathbf X}^{\text{CRPT}}$, $\mathbf R_{L_2} = [\mathbf r_{L_2}^{(1)}, \mathbf r_{L_2}^{(2)}] \in \mathbb R^{10 \times 2}$, and the $K=2$ $L_1$-principal components of $\tilde{\mathbf X}^{\text{CRPT}}$, $\mathbf R_{L_1} = [\mathbf r_{L_1}^{(1)}, \mathbf r_{L_1}^{(2)}] \in \mathbb R^{10 \times 2}$.
In Fig. \ref{fig:DOA}, we plot the standard $L_2$ MUSIC spectrum \cite{schmidt}
\begin{align}
P(\theta) \overset{\triangle}{=} \frac{1}{ \tilde{\mathbf s}_{\theta}^T (\mathbf I_{2D}- \mathbf R_{L_2}  \mathbf R_{L_2}^T) \tilde{\mathbf s}_{\theta}}, ~\theta \in \left( -\frac{\pi}{2}, \frac{\pi}{2}\right),
\end{align}
where  $ \tilde{\mathbf s}_{\theta} = [ \text{Re}\{{\mathbf s}_{\theta}\}^T, ~ \text{Im}\{{\mathbf s}_{\theta}\}^T ]^T$, as well as what we may call ``$L_1$ MUSIC spectrum'' with $\mathbf R_{L_1}$ in  place of $\mathbf R_{L_2}$.
 It is interesting to observe how $L_1$ MUSIC (in contrast to $L_2$ MUSIC) does not respond to the one-out-of-ten outlying jammer value in the data set and shows only the directions of the two actual nominal signals.

\subsection*{Experiment 4 - Image Reconstruction}

Consider  the ``clean'' $ 100 \times 64 $  gray-scale image $\mathbf A \in \{0, 1, \ldots, 255\}^{100 \times 64}$  of Fig.~\ref{fig:alexis1}.
We assume that $\mathbf A$ is not available and instead we have a data set of  $N=10$ corrupted/occluded versions of $\mathbf A$, say  $\mathbf A_{1}, \mathbf A_{2}, \ldots, \mathbf A_{10}$.
Each corrupted instance $\mathbf A_i$, $i=1, \ldots, 10$, is created by partitioning the original 
image $\mathbf A$  into sixteen tiles of size $25 \times 16$ and replacing three arbitrarily selected tiles by $25 \times 16$ grayscale-noise patches as seen, for example, in Fig.~\ref{fig:alexis2}.
 
 The $10$ corrupted instances are vectorized  to form the  data matrix
\begin{align}
\mathbf M= [ \text{vec}(\mathbf A_1), \ldots, \text{vec}(\mathbf A_{10}) ] \in \{0, \ldots, 255\}^{6400 \times 10}.
\end{align}
 Next, we  ``condense'' $\mathbf M$ to a  rank-$2$ representation by both $L_2$- and $L_1$-subspace projection,
\begin{align}
\hat{\mathbf  M}_{L_{2/1}}= \mathbf R_{L_{2/1}} \mathbf R_{L_{2/1}}^T \mathbf M,
\end{align}
where $\mathbf R_{L_{2/1}} \in \mathbb R^{6400 \times 2}$ consists of the $K=2$ $L_2$ or $L_1$, accordingly, principal components of $\mathbf M$.
In Fig.~\ref{fig:alexis3}  we show the projection of the corrupted image of Fig.~\ref{fig:alexis2} onto the  $L_2$-derived rank-$2$ subspace  (maximum-$L_2$-projection reconstruction).
In Fig.~\ref{fig:alexis4},   we show the projection of the same image onto the $L_1$-derived rank-$2$ subspace (maximum-$L_1$-projection reconstruction).
Figs. \ref{fig:alexis3} and (d) offer a perceptual (visual) interpretation of the difference between $L_2$ and $L_1$-subspace rank reduction. It is apparent that maximum-$L_1$-projection reconstruction offers a much clearer image representation of $\mathbf A$ than  maximum-$L_2$-projection reconstruction. This is another result that highlights the resistance of $L_1$-principal subspaces against outlying data corruption.

\section{Conclusions}

We presented for the first time in the literature optimal (exact) algorithms for the calculation of maximum-$L_1$-projection subspaces of data sets with complexity polynomial in the sample size (and exponent equal to the data dimension).
It may be possible in the future to develop an $L_1$ principal-component-analysis (PCA) line of research that parallels the enormously rewarding $L_2$ PCA/feature-extraction developments.
When $L_1$ subspaces are calculated on nominal ``clean" training data, they differ little --arguably-- from their $L_2$-subspace counterparts in least-squares fit.
When, however, subspaces are calculated from data sets with possible erroneous, out-of-line, ``outlier" entries, then $L_1$ subspace calculation offers significant robustness/resistance to the presence of inappropriate data values.\\
\mbox{}\\

\begin{center}
\bf{ACKNOWLEDGEMENT}\\
\end{center}
The authors would like to thank the Associate Editor and the four %
anonymous reviewers for their comments and suggestions that helped %
improve this manuscript significantly, both in presentation and content.

\newpage
\renewcommand{\arraystretch}{1}
 
\clearpage

\clearpage

\renewcommand{\arraystretch}{1.2}

\begin{figure} 
\centering
\begin{subfigure}[b]{0.6\textwidth}
\hrule
\vspace{0.5mm}
\hrule
\vspace{1.8mm}
\noindent {\bf The Optimal $L_1$-Principal-Component Algorithm}
\vspace{0.8mm}
\hrule
\noindent
\vspace{1.2mm}\\
\begin{tabular}{l}
~{\bf Input:} $\mathbf X_{D \times N}$ data matrix  
\end{tabular}\\
\begin{tabular}{r   l}
~1: &  $\left( \mathbf U_{N \times d}, \mathbf \Sigma_{d \times d}, \mathbf V_{d\times d}\right) \leftarrow \mathrm{svd} (\mathbf X^T)$\\
~2: &  $\mathbf Q_{N \times d} \leftarrow \mathbf {U \Sigma}$\\
~3: &  $\mathbf B_{N \times P_1} \leftarrow \mathrm{compute\_candidates} (\mathbf Q)$, $\mathcal P \leftarrow \{1, 2, \ldots, P_1 \}$\\
~4: &  $z_{\text{opt}} \leftarrow  {\arg \max}_{z \in \mathcal P}~  \| \mathbf X \mathbf B_{:,z} \|_2$\\
~5: &  $\mathbf b_{\text{opt}} \leftarrow \mathbf B_{:, z_{\text{opt}}}$\\
\end{tabular}\\
\begin{tabular}{l}
~\noindent {\bf Output:} $\mathbf r_{L_1} \leftarrow {\mathbf X \mathbf b_{\text{opt}}}/{\|\mathbf X \mathbf b_{\text{opt}}\|_2}$ 
\end{tabular}
 \vspace{0mm}\\
\hrule
\vspace{1.2mm}
\noindent {Function \emph{compute\_candidates}}
\vspace{0.8mm}
\hrule
\noindent
\vspace{1.2mm}\\
\begin{tabular}{l}
 ~{\bf Input:} $\mathbf Q_{N \times m}$
 \end{tabular}\\
 \begin{tabular}{r l}
1: & if $m > 2$,~  $i \leftarrow 0$\\
2: & ~~~~for $\mathcal I \subset \{1, 2, \ldots, N \}$ s.t. $| \mathcal I|=m-1$, $i \leftarrow i+1$,\\
3: & ~~~~~~~~$\bar{\mathbf Q}_{(m-1) \times m} \leftarrow \mathbf Q_{\mathcal I, :}$\\
4: & ~~~~~~~~$\mathbf c_{m \times 1} \leftarrow \mathrm{null}(\bar{\mathbf Q})$, $\mathbf c \leftarrow \mathrm{sgn}(c_m) \mathbf c$\\
5: & ~~~~~~~~$\mathbf B_{:,i} \leftarrow \mathrm{sgn}(\mathbf Q \mathbf c)$\\
6: & ~~~~~~~~for $j =1 : m-1$,\\
7: & ~~~~~~~~~~~~$ \mathbf c_{(m-1) \times 1} \leftarrow   \mathrm{null}(\bar{\mathbf Q}_{:/j,1:m-1}) $, $\mathbf c \leftarrow  \mathrm{sgn}(c_{m-1}) \mathbf c$\\
8: & ~~~~~~~~~~~~$\mathbf B_{\mathcal I(j), i} \leftarrow \mathrm{sgn}(\bar{\mathbf Q}_{j,1:m-1} \mathbf c)$\\
9: & ~~~~$\mathbf B \leftarrow [\mathbf B, \mathrm{compute\_candidates}(\mathbf Q_{:,1:m-2})]$\\
10: & elseif $m=2$,\\
11: & ~~~~for $i =1 : N$,\\
12: & ~~~~~~~~$\mathbf c_{2 \times 1} \leftarrow    \mathrm{null}( \mathbf Q_{i, :})$, $\mathbf c \leftarrow  \mathrm{sgn}(c_2) \mathbf c$\\
13: & ~~~~~~~~$\mathbf B_{:,i} \leftarrow \mathrm{sgn}(\mathbf Q \mathbf c)$, $\mathbf B_{i,i} \leftarrow \mathrm{sgn}({\mathbf Q}_{i,1})$ \\
14: & else, $\mathbf B \leftarrow \mathrm{sgn}(\mathbf Q)$\\
\end{tabular}\\
\begin{tabular}{l}
~{\bf Output:} $\mathbf B$ 
  \end{tabular}
\vspace{1mm}
\hrule
\vspace{0.5mm}
\hrule

 \vspace{4mm}

\end{subfigure}
\caption{The optimal ${\mathcal O}(N^d)$ algorithm for the computation of the maximum $L_1$-projection component of a rank-$d$ data matrix $\mathbf X_{D \times N}$ of $N$ samples of dimension $D$ (space complexity ${\mathcal O}(N)$; parallelizable computation of
columns of $\mathbf B$).}
\label{fig:algo}

\end{figure}

\begin{figure} 
\centering
\begin{subfigure}[b]{0.6\textwidth}
\hrule
\vspace{0.5mm}
\hrule
\vspace{1.8mm}
\noindent {\bf The Optimal $L_1$-Principal-Subspace Algorithm ($K>1$)}
\vspace{0.8mm}
\hrule
\noindent
\vspace{1.2mm}\\
\begin{tabular}{l}
~{\bf Input:} $\mathbf X_{D \times N}$ data matrix, subspace dimensionality $K$  
\end{tabular}\\
\begin{tabular}{r   l}
~1: &  $\left( \mathbf U_{N \times d}, \mathbf \Sigma_{d \times d}, \mathbf V_{d\times d}\right) \leftarrow \mathrm{svd} (\mathbf X^T)$\\
~2: &  $\mathbf Q_{N \times d} \leftarrow \mathbf {U \Sigma}$\\
~3: &  $\mathbf B_{N \times P_1} \leftarrow \mathrm{compute\_candidates} (\mathbf Q)$, $\mathcal P \leftarrow \{1, 2, \ldots, P_1 \}$\\
~4: &  $\mathbf z_{\text{opt}} \leftarrow  {\arg \max}_{\mathbf z \in \mathcal P^{K}, z_1 \leq z_2 \leq \ldots \leq z_K }~ \| \mathbf X \mathbf B_{:, \mathbf z} \|_*$\\
~5: &  $\mathbf B_{\text{opt}} \leftarrow \mathbf B_{:, \mathbf z_{\text{opt}}}$\\
~6: &  $\left( \mathbf U_{D \times K}, \mathbf \Sigma_{K \times K}, \mathbf V_{K\times K}\right) \leftarrow \mathrm{svd} (\mathbf X \mathbf B_{\text{opt}})$\\
 \end{tabular}\\
\begin{tabular}{l}
~\noindent {\bf Output:}  $\mathbf R_{L_1} \leftarrow \mathbf U \mathbf V^T$
\end{tabular}
\vspace{1mm}
\hrule
\vspace{0.5mm}
\hrule

 \vspace{4mm}

\end{subfigure}
\caption{The optimal ${\mathcal O}(N^{dK - K+1})$ algorithm for the computation of the  $K$-dimensional $L_1$-principal subspace of a rank-$d$ data matrix $\mathbf X_{D \times N}$ of $N$ samples of dimension $D$ (function \emph{compute\_candidates} in Fig. \ref{fig:algo}).}
\label{fig:algo_multi}

\end{figure}

\clearpage
 \renewcommand*{\arraystretch}{1.5}

\begin{table}
\begin{center}
\begin{tabular}{|c||*{6}{c|}}\hline
\backslashbox{{\hspace{0.2cm}$d$}}{\vspace{-0.2cm}{\hspace{-0.2cm}$N$}}
& ~~ 4 ~~ & ~~ 6 ~~ & ~~ 8 ~~ & ~~ 10 ~~ & ~~ 12 ~~ & ~~ 14 ~~ \\\hline\hline
3 & 0.0172 & 0.0406 & 0.0920 & 0.1966  & 0.3900 & 0.7160 \\\hline
4 & - &  0.0624  & 0.3526 & 1.4212 & 4.5178 & 11.8686\\\hline
5 & - & 0.1014 & 0.8471 & 5.4944 & 26.3361 & 99.4600\\\hline
6 & - & - & 1.2308 & 12.2289 & 87.1546 & 471.2275\\\hline
\end{tabular}
\end{center}
\label{tb:comp}
\caption{Average CPU time in seconds  expended by an Intel$^{{\tiny \textregistered}}$ Core$^{{\tiny \texttrademark}}$ i5 Processor (at 3.40 GHz) to find
the $(K=2)$-dimensional $L_1$-principal subspace of a full-rank $d \times N$ data matrix ($d<N$) by executing serially in Matlab$^{{\tiny \textregistered}}$ R2012a the algorithm of Fig. \ref{fig:algo_multi}.}
\end{table}

\clearpage

\begin{figure*} 
\centering
\begin{subfigure}[]{0.49\textwidth}
\centering
\includegraphics[width=\textwidth]{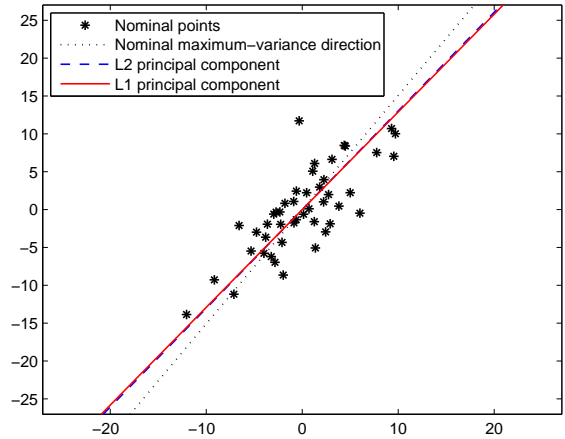}
\caption{}
\label{fig:dr1}
\end{subfigure}%
\\
\begin{subfigure}[]{0.49\textwidth}
\centering
\includegraphics[width=\textwidth]{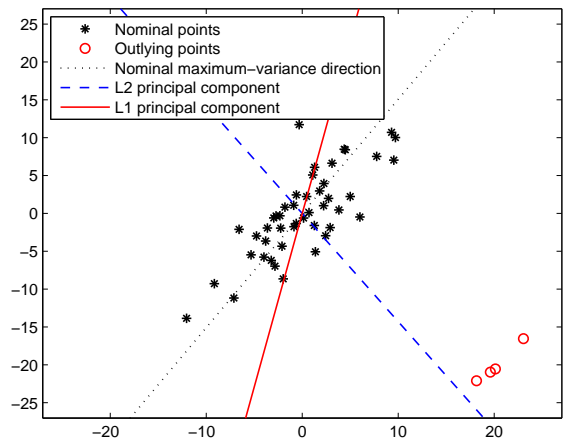}
\caption{ }
\label{fig:dr2}
\end{subfigure}

\begin{subfigure}[]{0.49\textwidth}
\centering
\includegraphics[width=\textwidth]{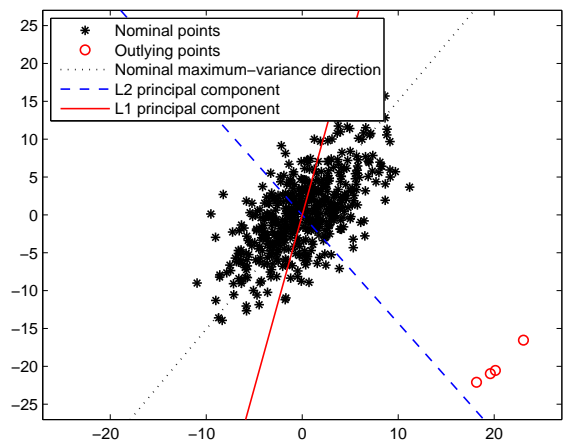}
\caption{}
\label{fig:dr3}
\end{subfigure}
\caption{
(a) Training data matrix $\mathbf X_{2 \times 50}$ with its $L_1$ and $L_2$ principal components ($K=1$).
(b) Training data matrix $\mathbf X_{2 \times 50}$ corrupted by four additional outlier points in bottom right with recalculated $L_1$ and $L_2$ principal components.
(c) Evaluation data set of $1000$ nominal points against the outlier infected (Fig. \ref{fig:dr2}) $L_1$ and $L_2$ principal components. For reference, in all figures we plot along the ideal maximum-variance direction of the nominal-data distribution
(dominant eigenvector of the true nominal-data autocovariance matrix).}
\label{fig:DR}
\end{figure*}

\clearpage

\begin{figure*}
\centering
\includegraphics[width=0.8\textwidth]{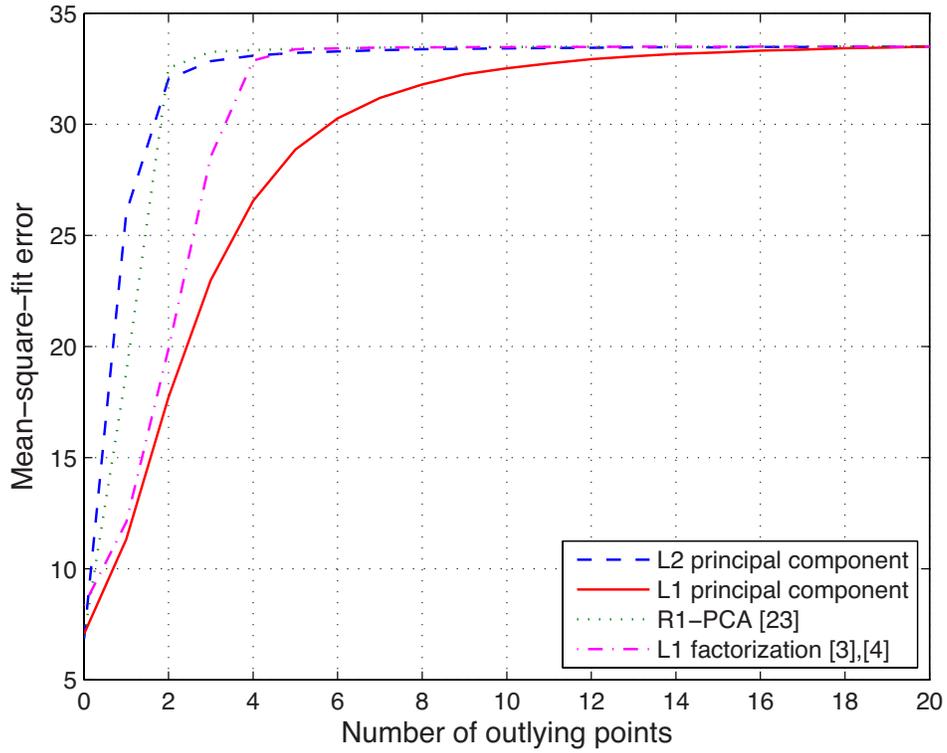}
\caption{Mean-square-fit error of $(D=2)$-dimensional data when projected onto the direction ($K=1$) of the $L_2$-principal component, the $L_1$-principal component, the $R_1$-principal component \cite{Ding2006}, and $L_1$-factorization \cite{Ke2003}, \cite{Ke2005}.}
\label{fig:linefitting_mse}
\end{figure*}

\clearpage

\begin{figure*}
\centering
\begin{subfigure}[]{.6\textwidth}
\centering
\includegraphics[width=\textwidth]{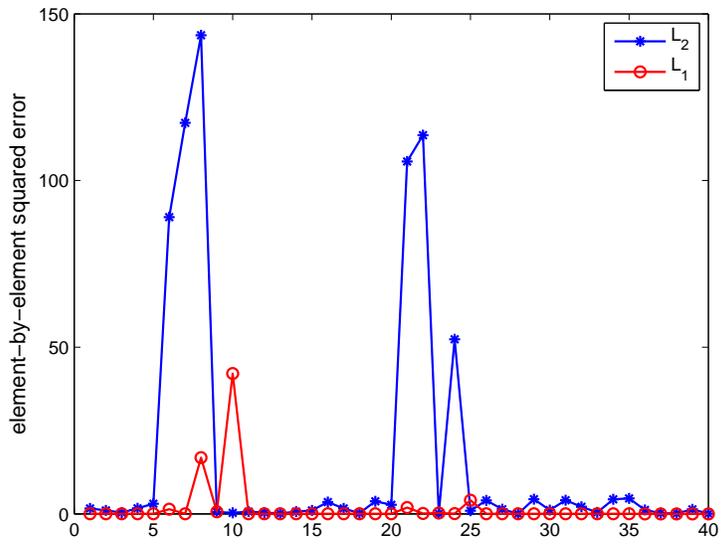}
\caption{}
\label{fig:error1}
\end{subfigure}\\
~
\begin{subfigure}[]{.6\textwidth}
\centering
\includegraphics[width=\textwidth]{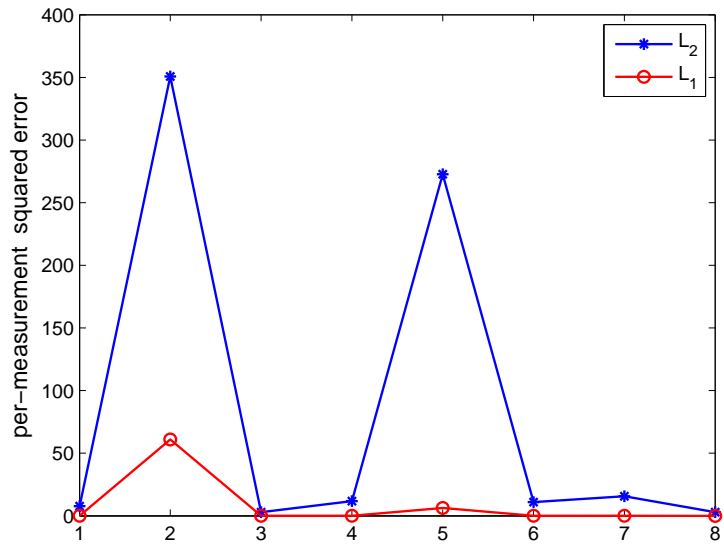}
\caption{ }
\label{fig:error2}
\end{subfigure}
\caption{(a) Element-by-element and (b) per-measurement  square restoration error.}
\label{fig:error}
\end{figure*}

\clearpage

\begin{figure*}
\centering
\includegraphics[width=0.8\textwidth]{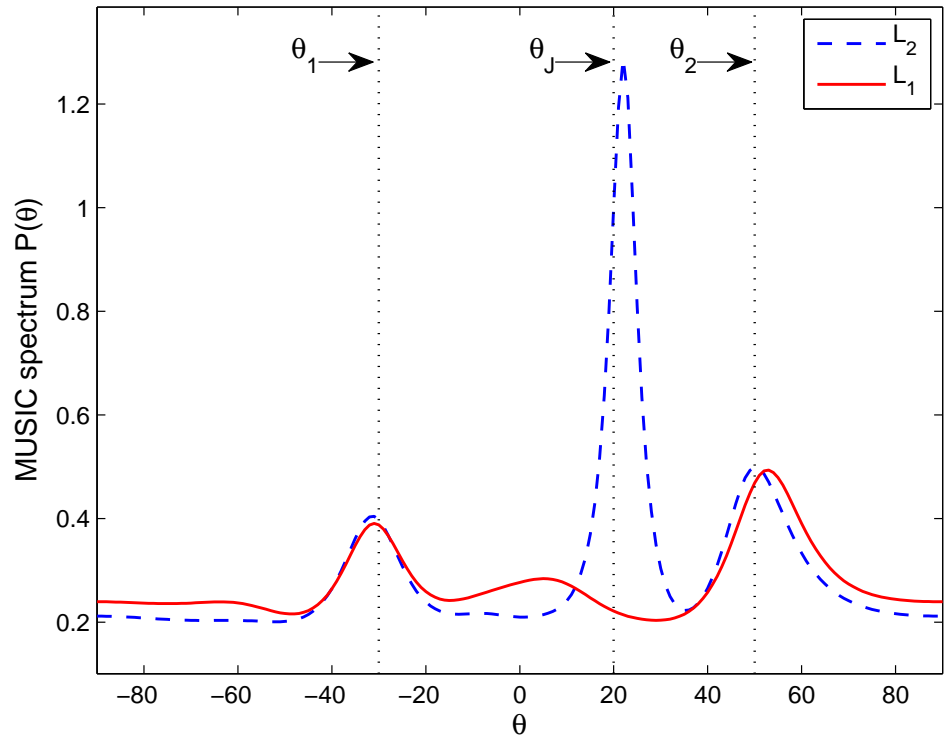}
\caption{MUSIC power spectrum with $K=2$ $L_2$ or $L_1$ calculated principal components (data set of $N=10$ measurements with signals at $\theta_1=-30^\circ$ and $\theta_2 = 50^\circ$ of which one measurement is additive-jammer corrupted with $\theta_J=20^\circ$; $\text{SNR}_1 = 2$dB; $\text{SNR}_2 = \text{SNR}_J = 3$dB).}
\label{fig:DOA}
\end{figure*}

\clearpage

\begin{figure*} 
\centering
\begin{subfigure}[]{0.15\textwidth}
\centering
\includegraphics[ width=0.8\textwidth]{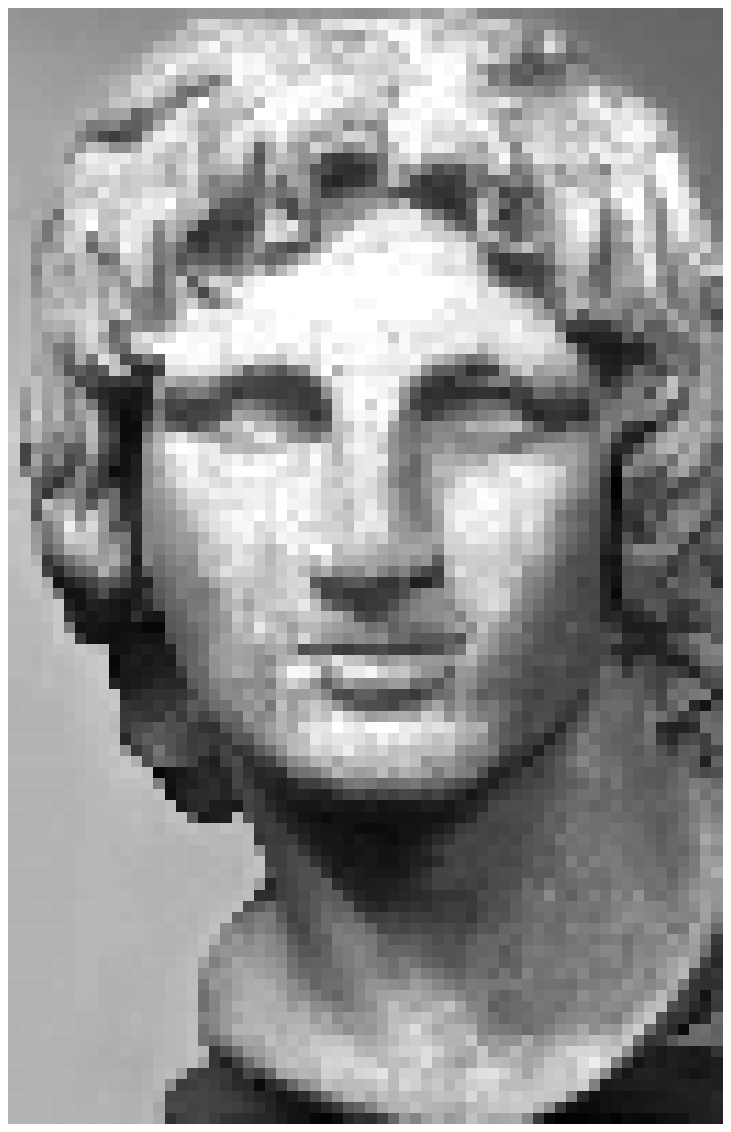}
\caption{}
\label{fig:alexis1}
\end{subfigure}%
~ 
\begin{subfigure}[]{0.15\textwidth}
\centering
\includegraphics [width=0.8\textwidth]{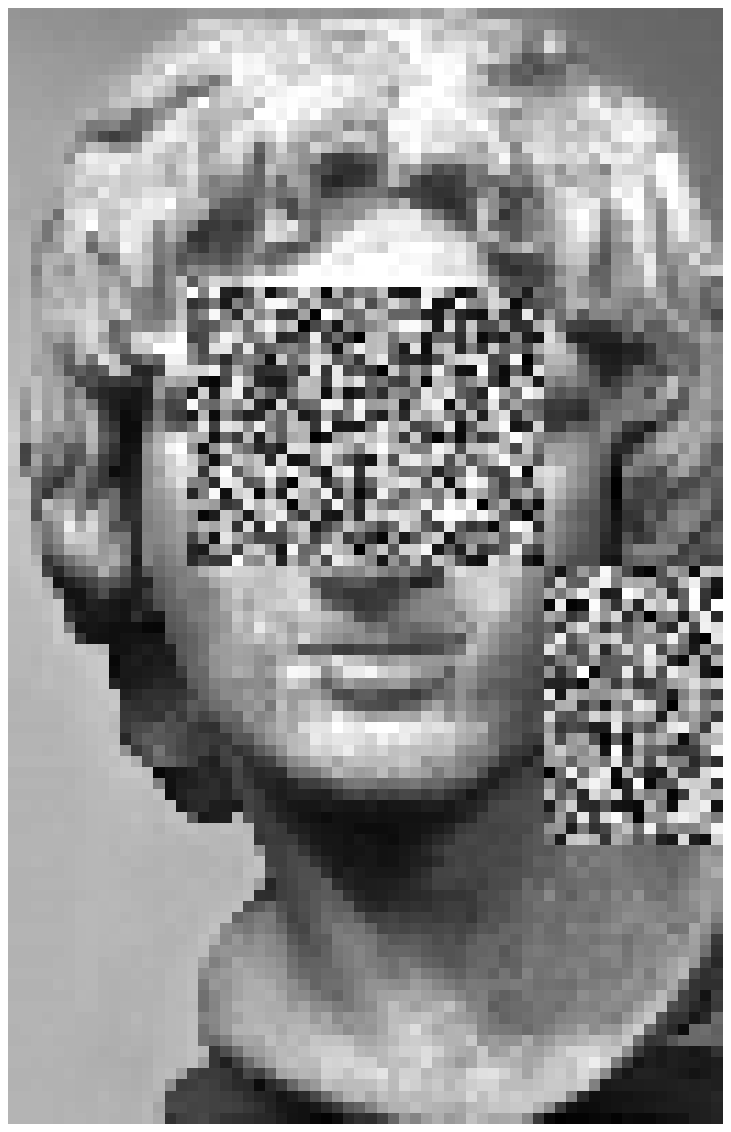}
\caption{ }
\label{fig:alexis2}  
\end{subfigure}
~
\begin{subfigure}[]{0.15\textwidth}
\centering
\includegraphics  [width=0.8\textwidth]{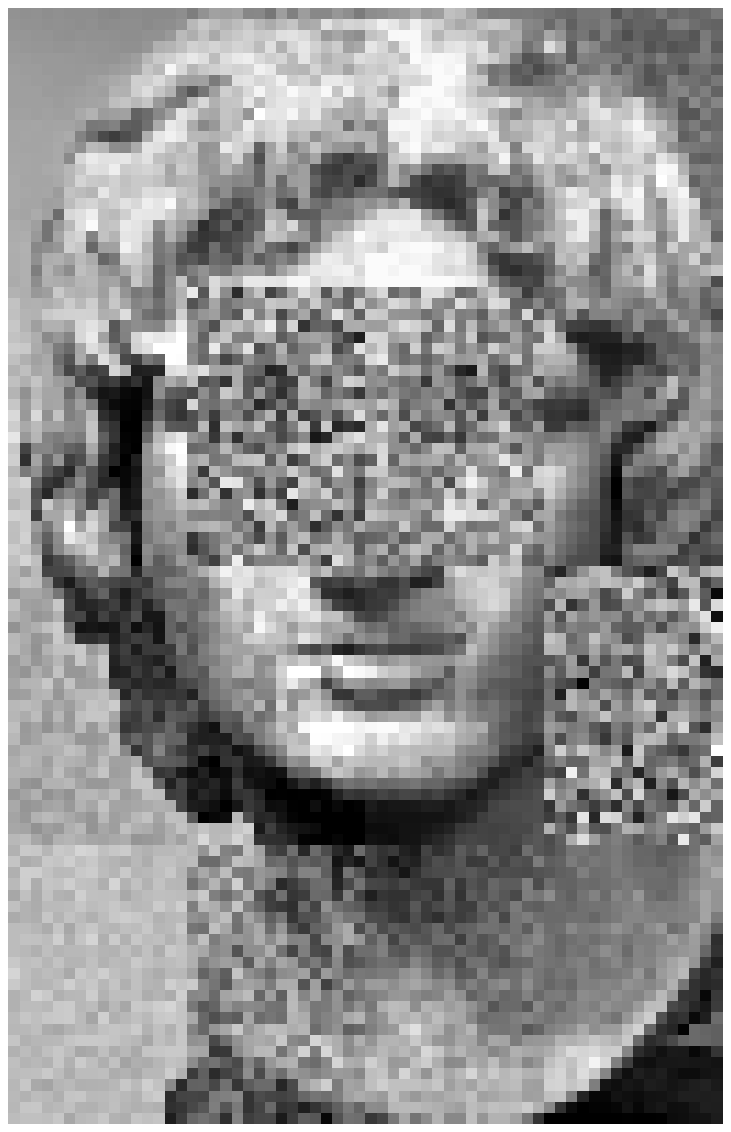}
\caption{}
\label{fig:alexis3}
\end{subfigure}%
~ 
\begin{subfigure}[]{0.15\textwidth}
\centering
\includegraphics [ width=0.8\textwidth]{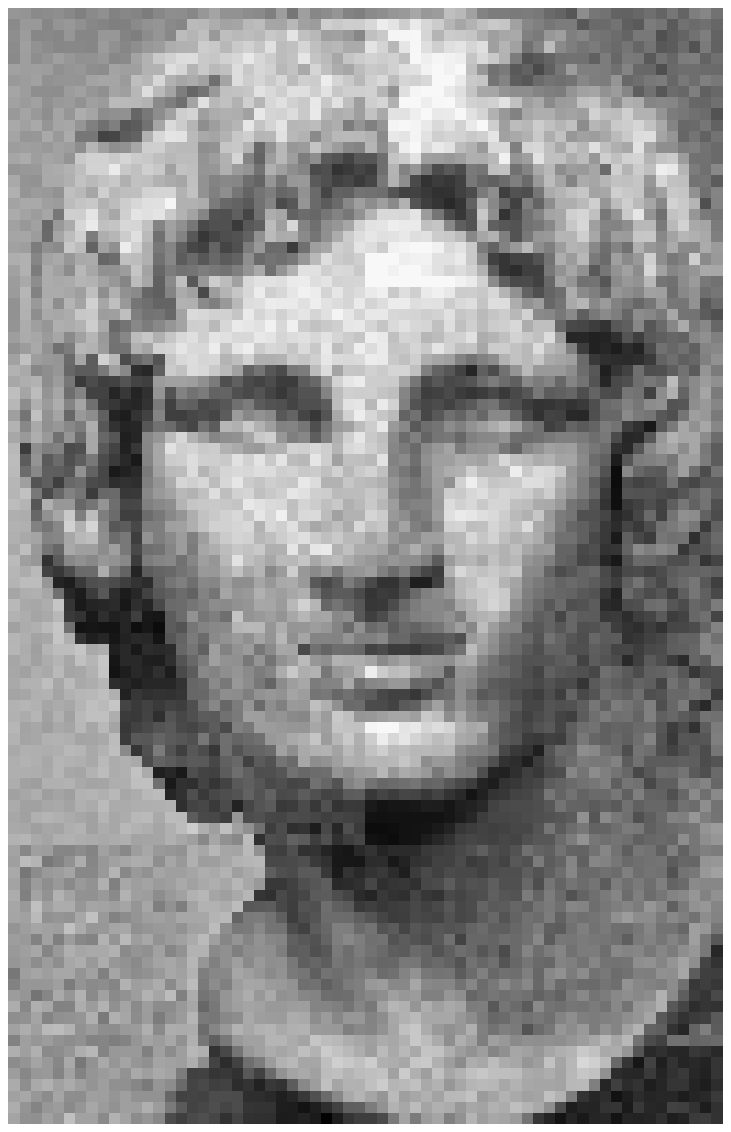}
\caption{ }
\label{fig:alexis4}  
\end{subfigure}
\caption{(a) Original image $\mathbf A \in \{ 0, 1, \ldots, 255 \}^{100 \times 64}$.
(b) An ``occluded'' instance of $\mathbf A$.
(c) Maximum-$L_2$-projection reconstructed image, and (d)  maximum-$L_1$-projection reconstructed image.}
\end{figure*}

\clearpage 

\end{document}